\documentclass[preprint,12pt,authoryear]{aastex}
\usepackage{amssymb}
\usepackage{natbib}
\usepackage{multirow}
\usepackage{amsmath}
\usepackage{graphicx}
\usepackage{longtable}
\slugcomment{\emph{Icarus}, in press, 24 June 2010}

\shorttitle{Chemical Constraints on Jupiter's Deep Water Abundance}

\shortauthors{Visscher, Moses, \& Saslow}
\begin{document}

\title{The Deep Water Abundance on Jupiter: New Constraints from Thermochemical Kinetics and Diffusion Modeling}

\author{Channon Visscher$^{1}$, Julianne I. Moses$^{2}$, and Sarah A. Saslow$^{3}$}

\affil{$^{1}$Lunar and Planetary Institute, USRA, Houston, Texas 77058-1113} \affil{$^{2}$Space Science Institute, Seabrook, Texas 77586}

\affil{$^{3}$University of Maryland, College Park, Maryland 20742}

\email{visscher@lpi.usra.edu, jmoses@SpaceScience.org}

\begin{abstract}
We have developed a one-dimensional thermochemical kinetics and diffusion model for Jupiter's atmosphere that accurately describes the transition from the thermochemical regime in the deep
troposphere (where chemical equilibrium is established) to the quenched regime in the upper troposphere (where chemical equilibrium is disrupted).  The model is used to calculate chemical
abundances of tropospheric constituents and to identify important chemical pathways for CO-CH$_4$ interconversion in hydrogen-dominated atmospheres. In particular, the observed mole fraction and
chemical behavior of CO is used to indirectly constrain the Jovian water inventory.  Our model can reproduce the observed tropospheric CO abundance provided that the water mole fraction lies in
the range $(0.25 - 6.0)\times10^{-3}$ in Jupiter's deep troposphere, corresponding to an enrichment of 0.3 to 7.3 times the protosolar abundance (assumed to be
H$_{2}$O/H$_{2}=9.61\times10^{-4}$).  Our results suggest that Jupiter's oxygen enrichment is roughly similar to that for carbon, nitrogen, and other heavy elements, and we conclude that
formation scenarios that require very large ($>\, $8 times solar) enrichments in water can be ruled out. We also evaluate and refine the simple time-constant arguments currently used to predict
the quenched CO abundance on Jupiter, other giant planets, and brown dwarfs.
\end{abstract}


\section{Introduction}\label{s Introduction}

The water abundance in Jupiter's deep atmosphere provides important clues for solar system formation and evolution and reveals conditions in the solar nebula at the time of giant-planet
formation \citep[e.g.,][]{lunine2004}.  In addition, the planetary water inventory has important implications for the cloud structure, energy balance, thermal structure, and chemistry of the
Jovian troposphere.  Unfortunately, the deep water abundance is difficult to obtain by remote sensing methods because H$_{2}$O is expected to condense near the $\sim$5 bar level in Jupiter's
cold upper troposphere \citep[e.g., see][]{taylor2004}, and clouds and other opacity sources limit the depth to which infrared and other wavelength radiation can penetrate. The only
\textit{in-situ} measurement of the Jovian water abundance, by the \textit{Galileo} Probe Mass Spectrometer (GPMS), pertains to a meteorologically anomalous ``hot-spot'' region characterized by
low cloud opacity, low mixing ratios of condensable species, high thermal emission, and a water abundance which increased with depth
\citep[e.g.,][]{orton1996,orton1998,niemann1998,ragent1998,sromovsky1998,wong2004}. Thus, it is unclear whether the probe descended deep enough to sample the well-mixed water abundance below the
cloud base, and the \textit{Galileo} probe value of H$_{2}$O/H$_{2}=4.9\pm1.6\times10^{-4}$ at the 19-bar level \citep{wong2004} is generally considered to be a lower limit for Jupiter's O/H
inventory \citep[e.g.,][]{roosserote2004,taylor2004,wong2004,depater2005}. For this reason, chemical models must be used to determine the deep water abundance in Jupiter's atmosphere until
further measurements become available, such as microwave observations from the \textit{Juno} mission or future deep-atmosphere probes \citep[e.g.,][]{bolton2006,atreya2004}.

Several investigators have used clever methods to estimate the deep H$_{2}$O abundance of the giant planets by considering the observed tropospheric abundance of CO and other trace constituents
and by investigating how H$_{2}$O chemistry and atmospheric transport can influence the abundance of these trace species
\citep[e.g.,][]{prinn1977,fegley1985apj,fegley1988,lodders1994,lodders2002,bezard2002,visscher2005}.  However, \citet{prinn1977}, and all subsequent modelers who used their kinetic schemes, were
limited by a lack of key chemical kinetics data, and some of the initial kinetic assumptions have been shown to be incorrect \citep{dean1987,yung1988,griffith1999,bezard2002,cooper2006}.  In
addition, some of the transport time-scale arguments in the earlier works have been shown to have inappropriate assumptions \citep{smith1998,bezard2002}. These problems offset each other to an
extent, such that models using earlier assumptions yield reasonable results \citep[cf.][]{visscher2005,bezard2002}. \citet{fegley1988} found that H$_{2}$O/H$_{2}= (0.46-5.8)\times10^{-3}$ is
consistent with CO kinetics and atmospheric mixing on Jupiter, whereas \citet{bezard2002} derived H$_{2}$O/H$_{2}=(0.34-15.3)\times10^{-3}$ using revised kinetic and transport time-scale
parameters \citep{page1989,yung1988,smith1998} and improved CO observations.  Further refinement of this estimate was not possible in the B\'ezard et al.~analysis due to uncertainties in
reaction kinetics, convective mixing rates, and the back-of-the-envelope time-scale arguments used to derive the H$_{2}$O abundance.

Here we attempt to improve the determination of the deep water abundance in Jupiter's atmosphere by taking advantage of recent updates in thermodynamic parameters and reaction rate coefficients
and by using a numerical model to provide a more rigorous quantitative test of the simple kinetic vs. transport time-scale approach.  With a numerical model, we implicitly solve the continuity
equations for all tropospheric constituents, considering reaction kinetics and atmospheric transport. As a result, our model tracks the transition from the thermochemical regime in the deep
troposphere (where chemical equilibrium is established), to a quenched regime in the upper troposphere (where chemical equilibrium is disrupted).  In contrast to previous studies, we make no
\textit{a priori} selection of the reaction mechanism, nor of the rate-determining step for the chemical conversion of CO into CH$_{4}$. Instead, we input a full suite of chemical kinetic
reactions connecting the different relevant tropospheric species, and allow the dominant chemical pathways for the conversion of CO$\rightarrow$CH$_{4}$ to be identified from our model results.
Furthermore, we make no assumptions about the mixing length scale but instead model atmospheric transport via diffusion for an assumed eddy diffusion coefficient profile.  We explore the effects
of the tropospheric water abundance on the chemical behavior of CO and other oxidized carbon gases and use the observed CO abundance to indirectly constrain the water inventory in Jupiter's deep
atmosphere.

The paper is organized as follows.  We begin with a description of our chemical model and the CO chemical constraint in \S\ref{s Computational Approach}.  We present our model results in
\S\ref{s Model Results}, derive an estimate of Jupiter's deep water abundance using CO as an observational constraint, identify the dominant chemical pathways involved for
CO$\rightarrow$CH$_{4}$ conversion in Jupiter's troposphere, and discuss the chemical behavior of other oxygen-bearing carbon species. In \S\ref{s: Discussion} we discuss implications of our
results for constraining Jupiter's total oxygen inventory and planetary formation scenarios, and conclude with a summary in \S\ref{s: Summary}.

\section{Chemical Model}\label{s Computational Approach}
\subsection{Numerical Approach}\label{ss Numerical Approach}
We use the Caltech/JPL KINETICS code \citep{allen1981} to calculate the vertical distribution of atmospheric constituents in Jupiter's troposphere by solving the coupled one-dimensional
continuity equations as a function of time $t$ and altitude $z$ for each species:
\begin{equation}\label{eq:continuity}
\frac{\partial n_{i}}{\partial t}+\frac{\partial \Phi_{i}}{\partial z}=P_{i}-L_{i},
\end{equation}
where $n_{i}$ is the number density (cm$^{-3}$), $\Phi_{i}$ is the vertical flux (molecules cm$^{-2}$ s$^{-1}$), $P_{i}$ is the chemical production rate (molecules cm$^{-3}$ s$^{-1}$), and
$L_{i}$ is the chemical loss rate (molecules cm$^{-3}$ s$^{-1}$) of species $i$.

The continuity equations are solved using finite-difference techniques for 144 atmospheric levels, with a vertical resolution of at least twenty altitude levels per scale height.  Jupiter's
pressure-temperature profile is taken from \textit{Galileo} entry probe data \citep{seiff1998} from 17 bar to the 22 bar, 427.7 K level (i.e., the maximum depth achieved by the probe before its
destruction) and is extended to greater temperatures and pressures along an adiabat using the method described by \citet{fegley1985apj}.  A zero flux boundary condition is maintained at the top
(17.4 bar, 399 K) and bottom (12,650 bar, 2500 K) of the model, with thermochemical equilibrium being used to define the initial conditions. The relative abundances of the elements are thus
defined {\it a priori}, and no mass enters or leaves the system.  Although our model does not include rock-forming elements which may react with oxygen in Jupiter's deep atmosphere, we do
consider the partial removal of oxygen into rock (see \S\ref{ss Atmospheric Composition} below) based upon the approach of \citet{lodders2004apj} and \citet{visscher2005} so that our abundance
calculations involving oxygen are correct for higher altitudes \citep[$T<2000$ K; cf.~Fig.~39 in][]{fegley1994}. Model calculations are performed until successive iterations differ by no more
than 0.1\%.

\subsection{Eddy Diffusion Coefficient}\label{ss Eddy Diffusion Coefficient}
In the context of our one-dimensional model, transport is assumed to occur by eddy diffusion, characterized by a vertical eddy diffusion coefficient $K_{zz}$.  In the absence of a strong
magnetic field or rapid rotation, the eddy diffusion coefficient can be estimated from free-convection and mixing-length theories \citep{stone1976,gierasch1985}, using the scaling relationship
\begin{equation}\label{eq: keddy}
K_{zz}\approx wH \approx \left(\frac{Fk_{\textrm{B}}}{\rho m c_{p}}\right)^{1/3}H
\end{equation}
where $w$ is the characteristic vertical velocity over which convection operates (cm s$^{-1}$), $F$ is Jupiter's internal heat flux \citep[5.44 W m$^{-2}$,][]{hanel1981, pearl1991},
$k_{\textrm{B}}$ is the Boltzmann constant, $\rho$ is the atmospheric mass density (g cm$^{-3}$), $m$ is the atmospheric mean molecular mass (g), and $c_{p}$ is the atmospheric specific heat at
constant pressure (erg K$^{-1}$ g$^{-1}$).  The characteristic length scale over which mixing operates is assumed to be the atmospheric pressure scale height $H$, given by
\begin{equation}\label{eq: Hscale}
H=\frac{k_{\textrm{B}}T}{m g}
\end{equation}
where $g$ is the gravitational acceleration.  Eq.~(\ref{eq: keddy}) yields $K_{zz}\sim7\times 10^{8}$ cm$^{2}$ s$^{-1}$ throughout much of Jupiter's deep troposphere.

However, strong magnetic fields or rapid rotation can affect convection and alter $K_{zz}$ \citep[e.g.,][]{flasar1977,stevenson1979,fernando1991}.  Rotation reduces $K_{zz}$, whereas magnetic
fields can either increase or decrease $K_{zz}$ over the estimates from standard mixing-length theory \citep{stevenson1979}. Considering rotation alone, we use the equations developed by
\citet{flasar1977,flasar1978} to describe small-scale thermally driven convection in a rapidly rotating atmosphere.  As outlined by these authors, the characteristic vertical velocity $w$ for
turbulent convection under these conditions is
\begin{align}
w \ \approx & \ \left(-\alpha g S \right)^{1/2} \, H^2, & \quad {\rm for}\ Ro \gg | \sin \lambda | \label{eq:wnorot}\\
w \ \approx & \ \left(\frac{Ro}{\sin \lambda}\right)^2 \left(-\alpha g S \right)^{1/2} \, H^2, & \quad {\rm for}\ Ro \ll | \sin \lambda | \label{eq:wrot}
\end{align}
where $\alpha$ is the thermal expansion coefficient ($= 1/T$ for an ideal gas), $S$ is the static stability $\partial \theta / \partial r$ (K cm$^{-1}$), i.e., the radial gradient of the
potential temperature $\theta$ (and note that $S < 0$ for free convection, as we are assuming here), $\lambda$ is the planetocentric latitude, and $Ro$ is the Rossby number associated with the
convective flow:
\begin{equation}
Ro = \frac{(-\alpha g S)^{1/2}}{2\Omega}
\end{equation}
where $\Omega$ is the rotational angular velocity (s$^{-1}$).  Note that Eq.~({\ref{eq:wnorot}), which is valid for near-equatorial latitudes, is the same expression that is derived from
mixing-length theory in non-rotating systems (see Eq.~(\ref{eq: keddy}) above).  At non-equatorial latitudes, Eq.~({\ref{eq:wrot}) is appropriate, although it is convenient to replace $S$ with
an expression involving $F$, the internal heat flux --- a measured quantity.  From mixing-length theory,
\begin{equation}
w \approx \frac{-F}{\rho c_p H S} \label{eq:wmix}.
\end{equation}
After equating Eq.~(\ref{eq:wrot}) and Eq.~(\ref{eq:wmix}) and solving for $S$ and $Ro$ in terms of $F$, we find that
\begin{equation}
K_{zz} \approx \left[\frac{{k_{\textrm{B}}}F H^{1/3}}{\rho m c_{p} (4\Omega^{2} \sin^{2} \lambda)^{2/3}}\right]^{3/5} \quad \quad {\rm for}\  Ro \ll | \sin \lambda | \label{eq:Krot}
\end{equation}

Figure \ref{fig:Krotus} shows the solution to Eq.~(\ref{eq:Krot}) for non-equatorial latitudes over our model pressure range, although we should note that the condition $Ro \ll | \sin \lambda |$
begins to break down at high altitudes and low latitudes (e.g., $Ro = |\sin \lambda|$ for 20 bar at 3$^\circ$ latitude).  Note that these $K_{zz}$ values are smaller than those predicted by
standard mixing-length theory in a non-rotating atmosphere (see above); moreover, there is a very strong latitude dependence of $K_{zz}$ at any particular pressure. As noted by Flasar and
Gierasch (1977), such meridional variations in $K_{zz}$ could have some interesting consequences, including the production of meridional gradients in temperature and in the mixing ratios of
quenched disequilibrium species like CO.  However, these gradients may in turn drive meridional fluxes that act to homogenize the temperatures and abundances.  It is therefore difficult to
determine an appropriate $K_{zz}$ profile to adopt for our models.

The \textit{Galileo}\ probe entered the Jovian atmosphere in the Northern Equatorial Belt (NEB) near 6.5$^\circ$ N latitude, and the CO observations of \citep{bezard2002} were also centered in
the NEB near 9$^\circ$ N latitude.  At 9$^\circ$ latitude, Eq.~(\ref{eq:Krot}) implies $K_{zz}$ $\sim$ 4$\, \times\, 10^7$ cm$^2$ s$^{-1}$ at 2400 bar, increasing to 1$\, \times\, 10^8$ cm$^2$
s$^{-1}$ at 250 bar and 2.5$\, \times\, 10^8$ cm$^2$ s$^{-1}$ at 20 bar.  Because of uncertainties in the appropriate values of $K_{zz}$ in Jupiter's deep troposphere, we adopt an
altitude-independent $K_{zz}$ value of $1\times 10^{8}$ cm$^{2}$ s$^{-1}$ for our nominal model, but we also explore the effects of a range of altitude-independent $K_{zz}$ values from $4\times
10^{7}$ to $1\times 10^{9}$ cm$^{2}$ s$^{-1}$, as well as consider $K_{zz}$ values that vary with altitude using Eq.~(\ref{eq:Krot}).  We assume that this range encompasses all plausible values
for $K_{zz}$ near the quench level.

\subsection{Atmospheric Composition}\label{ss Atmospheric Composition}
Our model includes 108 hydrogen, carbon, nitrogen, oxygen, and phosphorus species that are subject to vertical transport and chemical production and loss.  The measured relative abundances of
these elements (see Table \ref{tab: gas abundances}) are used to define our initial conditions.  The observed abundances of sulfur (as H$_{2}$S) and the noble gases Ar, Kr, and Xe are shown in
Table \ref{tab: gas abundances} for comparison.  Atmospheric mixing ratios for He, CH$_{4}$ and NH$_{3}$ are taken from GPMS and helium interferometer measurements: He/H$_{2}=0.1574$
\citep{vonzahn1998,niemann1998}, and CH$_{4}$/H$_{2}=2.37 \times 10^{-3}$ and NH$_{3}$/H$_{2}=6.64\times10^{-4}$ \citep{wong2004}.  Our adopted phosphorus elemental abundance is based on a deep
phosphine mixing ratio of PH$_{3}$/H$_{2}=8\times 10^{-7}$ derived from \textit{Voyager} IRIS \citep{kunde1982,lellouch1989}, \textit{Galileo} NIMS \citep{irwin1998}, Kuiper Airborne Observatory
\citep{bjoraker1986} and ground-based \citep{bezard2002} 5-$\mu$m observations of Jupiter.

The \textit{Galileo} entry probe measured a tropospheric water abundance of H$_{2}$O/H$_{2}=4.9\times 10^{-4}$ near the 19-bar level \citep{wong2004}, which corresponds to a subsolar O/H ratio.
As was previously mentioned, it is unclear whether this value is representative of Jupiter's deep, well-mixed water inventory because the probe entered a meteorologically dry hot-spot region and
because the H$_{2}$O abundance increased nearly tenfold over the previous measurement at the 11-bar level \citep[see][]{wong2004}.  Subsolar O/H values are also consistent with 5-$\mu$m
observations from the Kuiper Airborne Observatory (KAO) \citep{larson1975,bjoraker1986}, the \textit{Voyager} Infrared Interferometer Spectrometer (IRIS)
\citep{drossart1982,kunde1982,bjoraker1986,lellouch1989}, the \textit{Galileo} Near-Infrared Mapping Spectrometer (NIMS) \citep{irwin1998,roosserote1998}, and the Infrared Space Observatory
(ISO) \citep{encrenaz1996,roosserote1999,roosserote2004}. However, the interpretation of the observations is complicated by several factors: (1) these 5-$\mu$m observations are weighted by
thermal emission from the relatively dry and cloud-free hot-spot regions, (2) the observations are most sensitive to an altitude region in which water is expected to condense so that the H$_2$O
mole fraction will be varying strongly with altitude, and (3) the results are highly dependent on assumptions regarding cloud structure and extinction.  As is discussed by
\citet{roosserote2004}, the \textit{Galileo}\ NIMS observations are consistent overall with a deep O/H ratio of 1-2 times solar (see the discussion below on the solar ratio) and
\citet{roosserote2004} state that ``subsolar values of the O/H ratio cannot be reconciled with the analyzed data,'' but this conclusion is also model dependent. Due to the difficulties in
determining the O/H ratio from remote-sensing observations and due to the ambiguities in interpreting whether the probe data are representative of the whole planet, we treat the oxygen elemental
abundance as a free parameter in our model
--- indeed, the main free parameter that we are ultimately trying to constrain from our model-data comparisons --- and we consider a wide range of H$_{2}$O/H$_{2}$ ratios in order to explore the
effect of this parameter on tropospheric chemistry and the CO abundance.

Table \ref{tab: gas abundances} also contains the elemental enrichment factors relative to the assumed protosolar abundances.  Elemental abundances for the solar nebula (i.e., protosolar
abundances) are taken from \citet{lodders2009}.  These abundances are slightly different from present-day solar photospheric abundances because of gravitational settling of heavy elements in the
Sun \citep{lodders2003} and represent the bulk elemental composition of the Sun and solar nebula. As in previous studies \citep{lodders2004apj,visscher2005}, the protosolar H$_{2}$O/H$_{2}$
ratio is defined by taking the total oxygen abundance ($\Sigma$O/H$_{2}=1.21\times10^{-3}$) and subtracting the portion that forms rock (O$_{\textrm{rock}}$):
\begin{equation}\label{eq: oxygen mass balance}
\textrm{O}_{\textrm{H}_{2}\textrm{O}}=\Sigma \textrm{O} - \textrm{O}_{\textrm{rock}}.
\end{equation}
In a gas with a protosolar composition \citep{lodders2009}, the formation of rock effectively removes $\sim$20\% of the total oxygen inventory.   Throughout the following, we thus define the
water enrichment factor ($E_{\textrm{H}_{2}\textrm{O}}$) as
\begin{equation}
E_{\textrm{H}_{2}\textrm{O}}=(\textrm{H}_{2}\textrm{O}/\textrm{H}_{2})_{\textrm{Jupiter}}/(\textrm{H}_{2}\textrm{O}/\textrm{H}_{2})_{\textrm{solar}},
\end{equation}
in which enrichments over ``solar'' refer to a protosolar abundance of H$_{2}$O/H$_{2}$ $=9.61\times10^{-4}$.  Using this definition, the \textit{Galileo} probe measured a water abundance
equivalent to 0.51x solar ($E_{\textrm{H}_{2}\textrm{O}}=0.51$) in Jupiter's troposphere.

The relative abundance of H$_{2}$ (which comprises $86\%$ of the atmosphere) is calculated by difference using the observed mole fractions for He, CH$_{4}$, NH$_{3}$, PH$_{3}$, and varying
assumptions for the H$_{2}$O mole fraction. From the adopted abundances of these major compounds, the initial chemical composition for all 108 species in the model is calculated along Jupiter's
pressure-temperature profile by thermochemical equilibrium.  We use the NASA CEA (Chemical Equilibrium with Applications) code \citep{gordon1994} for our equilibrium calculations, utilizing
thermodynamic parameters from the compilations of \citet{gurvich1989,gurvich1991,gurvich1994,chase1998,burcat2005} and other literature sources. Figure \ref{fig:carbeq} shows our equilibrium
results for carbon chemistry in Jupiter's atmosphere, which are similar to the results of \citet[][see their Fig.~17]{fegley1994} with the exception of H$_{2}$CO and CH$_{2}$: here we use the
H$_{2}$CO enthalpy of formation from \citet{gurvich1989,gurvich1991,gurvich1994} instead of \citet{chase1998} based upon the recommendation of \citet{dasilva2006}, and we calculate the mole
fraction abundances of $^{1}$CH$_{2}$ and $^{3}$CH$_{2}$ individually using updated thermodynamic parameters from \citet{ruscic2005}.  From this initial equilibrium condition, we run the
kinetic-transport model to solve the continuity equations (described by Eq.~\ref{eq:continuity}) until steady state is achieved.

\subsection{Chemical Reactions}\label{ss: Chemical Reactions}
Approximately 1800 chemical reactions are included in our model.  Photolysis reactions are omitted because the tropospheric pressure levels under consideration ($\geq 17$ bars) are too deep for
ultraviolet photons to penetrate.  Our reaction list is based upon previous chemical models of giant-planet atmospheres
\citep[e.g.,][]{gladstone1996,moses1995a,moses1995b,moses2000a,moses2000b,moses2005} and includes numerous rate coefficient updates from combustion chemistry studies over the past two decades
\citep[e.g.,][]{baulch1992,baulch1994,baulch2005,smith1999,dean2000,korobeinichev2000,gardiner2000,miller2005}.

All of the reactions in the model are reversed using the principle of microscopic reversibility,
\begin{equation}\label{eq:reaction reversal}
K_{eq}=\frac{k_{f}}{k_{r}},
\end{equation}
where $K_{eq}$ is the equilibrium constant, $k_{f}$ is the rate coefficient for the forward reaction, and $k_{r}$ is the rate coefficient for the reverse reaction. The thermodynamic parameters
for calculating $K_{eq}$ are taken primarily from the compilations of \citet{gurvich1989,gurvich1991,gurvich1994}, \citet{chase1998}, and \citet{burcat2005}. We also adopt updated thermodynamic
data for several radical species (notably OH, CH$_{3}$, CH$_{2}$OH and CH$_{3}$O) from \citet{ruscic2002,ruscic2005}.  Equation (\ref{eq:reaction reversal}) is used to determine $k_{r}$ for
every reaction at each of the 144 atmospheric levels using the appropriate temperature-dependent values for $k_{f}$ and $K_{eq}$.  The outcome of this procedure is a reaction list which consists
of $\sim$900 forward reactions and $\sim$900 complimentary reverse reactions, which can be used to accurately reproduce equilibrium abundances based upon the selected thermodynamic parameters.
In other words, in the absence of disequilibrium effects such as photochemistry or atmospheric mixing, and given sufficient time to reach a steady state, our kinetic model results are
indistinguishable from those given by Gibbs energy minimization or mass-action, mass-balance thermodynamic-equilibrium calculations.

For three-body (termolecular) reactions, we assume that the forward reaction rate constant $k_{f}$ (cm$^{6}$ s$^{-1}$) obeys the expression
\begin{equation}\label{eq:termolecular rate coefficient}
k_{f}=\left(\frac{k_{0}}{1+\frac{k_{0}[\textrm{M}]}{k_{\infty}}}\right) F_{c}^{\beta},
\end{equation}
where $k_0$ is the low-pressure three-body limiting value (cm$^{6}$ s$^{-1}$), $k_{\infty}$ is the high-pressure limiting value (cm$^{3}$ s$^{-1}$), and $[\textrm{M}]$ is the total atmospheric
number density (cm$^{-3}$).  The exponent $\beta$ in Eq. (\ref{eq:termolecular rate coefficient}) is given by
\begin{equation}
\beta=\left(\frac{1}{1+\left[\log_{10}\left(\frac{k_{0}[\textrm{M}]}{k_{\infty}}\right)\right]^{2}}\right),
\end{equation}
and we assume $F_{c}\approx 0.6$ \citep{demore1992} except where specified in \citet{moses2005}.  Termolecular and unimolecular reactions in the model are also reversed using the method
described above (Eq. (\ref{eq:reaction reversal})).

\subsection{CO Chemical Constraint}\label{s: CO Chemical Constraint}
Carbon monoxide was first detected on Jupiter by \citet{beer1975} in the 5-$\mu$m window, with follow-up observations by \citet{beer1978}, \citet{larson1978}, and \citet{noll1988,noll1997}.
High-spectral-resolution observations that resolve the line shapes are needed to determine the origin of the CO, which can have either an internal source (i.e., from rapid mixing from the deep
troposphere) or an external source (e.g., from satellite or meteoroidal debris).  We are concerned solely with the internal source in this paper.  As such, we rely on the observations that have
most definitively resolved the CO vertical profile, those of \citet{bezard2002}.  From ground-based $\sim$4.7 $\mu$m observations with a very high spectral resolution of 0.045 cm$^{-1}$,
\citet{bezard2002} determined that both internal and external sources contribute to CO on Jupiter and concluded that the derived tropospheric mole fraction of $(1.0\pm0.2)\times10^{-9}$
represents the contribution from the internal source.

Our basic approach for constraining Jupiter's water inventory is similar to the method adopted by previous authors \citep[e.g.,][]{fegley1988,fegley1994,lodders1994,bezard2002,visscher2005}: the
observed abundance and chemical behavior of CO is used to estimate the H$_{2}$O abundance in Jupiter's atmosphere. Deep in the troposphere, CO is produced from water and methane via the net
thermochemical reaction
\begin{equation}\label{r: CO net thermochemical reaction}
\textrm{CH}_{4}+\textrm{H}_{2}\textrm{O} = \textrm{CO} + 3\textrm{H}_{2} \quad ,
\end{equation}
At high pressures and low temperatures, the reactants on the left-hand side of the reaction (CH$_{4}$, H$_{2}$O) are favored, whereas CO becomes more stable at low pressures and high
temperatures. On Jupiter, the temperature factor is more important, and the CO abundance is expected to increase with depth, although methane remains the dominant carbon-bearing gas throughout
the atmosphere. The equilibrium constant expression for reaction (\ref{r: CO net thermochemical reaction}) may be written as
\begin{equation}\label{K: CO net thermochemical reaction}
K_{\ref{r: CO net thermochemical reaction}}=\frac{[\textrm{CO}][\textrm{H}_{2}]^{3}}{[\textrm{CH}_{4}][\textrm{H}_{2}\textrm{O}]},
\end{equation}
where $K_{\ref{r: CO net thermochemical reaction}}$ is the equilibrium constant for reaction (\ref{r: CO net thermochemical reaction}) and  $[i]$ is the concentration (cm$^{-3}$) of each species
$i$. Rearranging this expression,
\begin{equation}\label{eq: XCO proportional}
[\textrm{CO}]=K_{\ref{r: CO net thermochemical reaction}}[\textrm{CH}_{4}][\textrm{H}_{2}\textrm{O}]/[\textrm{H}_{2}]^{3},
\end{equation}
shows that [CO]$\, \propto \, $[H$_{2}$O] at equilibrium for constant pressure and temperature conditions \citep[e.g.,][]{fegley1988,visscher2005}.  Thus, if thermochemical equilibrium holds,
the abundance of any one of these species in reaction (\ref{r: CO net thermochemical reaction}) at a given altitude (i.e., at a given pressure and temperature) can be readily calculated from the
observed abundances of the other three species at the same altitude in Jupiter's atmosphere.

However, the observed CO mole fraction of $X_{\textrm{CO}}=(1.0\pm0.2)\times10^{-9}$, reported by \citet{bezard2002} for the 6-bar level in Jupiter's atmosphere, is many orders of magnitude
higher than the CO abundance predicted under thermochemical equilibrium conditions for any plausible assumption about the deep water abundance on Jupiter and provides clear evidence of
disequilibrium processes at work in the troposphere \citep[e.g.,][]{prinn1977,barshay1978,fegley1985apj,fegley1988,fegley1994,bezard2002}. Carbon monoxide is not in equilibrium in Jupiter's
upper troposphere, and therefore Eq.~(\ref{eq: XCO proportional}) cannot be used to derive Jupiter's deep water inventory.

\citet{prinn1977} demonstrated that the observed CO abundance likely results from mixing from deeper atmospheric levels where CO is more abundant.  As parcels of gas rise in Jupiter's
atmosphere, CO is destroyed by conversion into CH$_{4}$ at a rate that falls off dramatically with decreasing temperatures.  The observable amount of CO in the upper atmosphere thus depends upon
the relative time scales of CO destruction kinetics (characterized by $t_{chem}$) and convective vertical mixing (characterized by $t_{mix}$) \citep[e.g., see][]{prinn1977,fegley1985apj}. Deep
in the troposphere, $t_{chem} < t_{mix}$ and thermochemical equilibrium is achieved because kinetic reactions dominate over atmospheric mixing.  At higher, colder altitudes, $t_{chem} >
t_{mix}$, and disequilibrium prevails because vertical mixing operates faster than reaction kinetics can attain equilibrium. At some intermediate ``quench'' level defined by $t_{chem}=t_{mix}$,
the CO mole fraction becomes quenched; above this level, the CO abundance remains fixed at the equilibrium mole fraction achieved at the quench level \citep{prinn1977}. As a result, the observed
CO mole fraction in Jupiter's upper atmosphere still serves as useful chemical probe for conditions in Jupiter's deep atmosphere \citep[e.g.,][]{fegley1983,fegley1994,bezard2002}, despite the
departure from equilibrium, provided that the kinetics of the CO $\rightarrow$ CH$_4$ conversion process is accurately known (to define $t_{chem}$) and the rate of atmospheric mixing can be
constrained (to define $t_{mix}$).

Although we go beyond the back-of-the-envelope $t_{chem}$-$t_{mix}$ approach with our kinetic-transport model, the underlying principle is the same.  We determine the range of water enrichments
that are consistent with the observed CO mole fraction for different plausible assumptions of the eddy diffusion coefficient and thereby indirectly constrain Jupiter's global water inventory.

\section{Model Results}\label{s Model Results}
\subsection{CO Profiles and Their Sensitivity to Model Free Parameters}\label{ss: Sensitivity}
The CO vertical profiles resulting from our thermochemical kinetics and diffusion model are presented in Fig.~\ref{fig: CO keddy} for various assumptions about the eddy diffusion coefficient and
for a single assumption about the deep water abundance (1x solar).  Shown for comparison is the observed CO mole fraction of $X_{\textrm{CO}}=(1.0\pm0.2)\times10^{-9}$ reported by
\citet{bezard2002} for the 6-bar level on Jupiter; note that we have assumed that this value remains constant down to at least the few tens of bar region.  The dashed gray line in Fig.~\ref{fig:
CO keddy} shows the predicted equilibrium abundance of CO (calculated using the NASA CEA code), which increases with depth. In the deep atmosphere ($P\gtrsim 600$ bars, $T\gtrsim 1100$ K), the
modeled CO abundance follows thermochemical equilibrium because the energy barriers for chemical reactions are easily overcome at these higher temperatures.  As a result, the characteristic
chemical time scale for interconversion between CO and CH$_{4}$ is much shorter than the time scale for vertical mixing ($t_{chem} \ll t_{mix}$), and equilibrium is readily achieved. However, as
atmospheric parcels are transported to higher, colder altitudes, the chemical kinetic timescale becomes longer (i.e., the reactions become slower) relative to the mixing time scale, and the CO
mole fraction can become quenched.  In Fig.~\ref{fig: CO keddy}, divergence from the equilibrium profile occurs where vertical convective mixing begins to drive the CO abundance out of
equilibrium and toward a constant quenched mole fraction. The transition region between the equilibrium and quenched regimes (where $t_{chem}\approx t_{mix}$) is not abrupt but occurs over a
range of altitudes approximately equal to one pressure scale height: for example, at $K_{zz}=1\times10^{8}$ cm$^{2}$ s$^{-1}$, the modeled CO abundance diverges from equilibrium near the
$640$-bar level and resumes a constant quenched profile above the $200$-bar level.  Our numerical models confirm the analytic prediction of \citet{prinn1977} that the CO mole fraction remains
fixed at altitudes above the quench region when the chemical reaction kinetics for CO$\, \rightarrow \, $CH$_{4}$ are extremely slow relative to the rate of vertical mixing ($t_{chem} \gg
t_{mix}$).

Figure \ref{fig: CO keddy} illustrates that the ``quenched'' upper tropospheric CO mole fraction depends on the strength of convective mixing.  For stronger convective mixing (i.e., larger
$K_{zz}$), CO is quenched deeper in the atmosphere where the CO mole fraction is larger.  Conversely, weaker mixing results in lower quenched CO mole fractions.  If the deep water enrichment is
1x solar, Fig.~\ref{fig: CO keddy} shows that the CO observations are best reproduced for $K_{zz}$ = $4\times10^{8}$ cm$^2$ s$^{-1}$, if we assume a constant $K_{zz}$ profile with altitude. If
the $K_{zz}$ profile varies with altitude, as with the model represented by a black dashed line in Fig.~\ref{fig: CO keddy}, the results are controlled by the $K_{zz}$ value near the quench
level.

The quenched CO mole fraction also depends on the assumed deep water abundance. Figure \ref{fig: CO water} shows vertical abundance profiles for CO in Jupiter's atmosphere for our assumed
nominal $K_{zz}$ value of $1\times10^{8}$ cm$^{2}$ s$^{-1}$ over a range of water enrichments, including the abundance measured by the \textit{Galileo} entry probe (0.51x solar).  The gray lines
indicate the equilibrium abundance of carbon monoxide for different water enrichments; divergence from the equilibrium profiles occurs where fast atmospheric mixing and slow reaction kinetics
(relative to one another) quench the CO mole fraction.  For any single $K_{zz}$ profile, the pressure level at which quenching occurs is roughly independent of the H$_{2}$O abundance, over the
range of water enrichments considered here.  Therefore, for all the models shown in Fig.~\ref{fig: CO water}, the quench level occurs at similar pressure and temperature conditions.  As
discussed above, for otherwise constant conditions (e.g., \textit{P}, \textit{T}, \textit{K}, [CH$_{4}$], [H$_{2}$]; see Eq. \ref{eq: XCO proportional}), the CO abundance is linearly
proportional to the H$_{2}$O abundance in Jupiter's deep atmosphere (see also Fig.~\ref{fig: CO sensitivity}).  The quenched CO mole fraction thus increases as the assumed deep water abundance
increases. For the assumption $K_{zz}=1\times10^{8}$ cm$^{2}$ s$^{-1}$, Fig.~\ref{fig: CO water} shows that the CO observations are best reproduced for a global water enrichment between 2-4
times solar.

Table \ref{tab: model results} and Fig.~\ref{fig: CO sensitivity} summarize the sensitivity of the quenched upper tropospheric CO mole fraction to variations in the two main free parameters of
our kinetic/transport model: the atmospheric convective mixing rate (characterized by $K_{zz}$) and the tropospheric water inventory (characterized by $E_{\textrm{H}_{2}\textrm{O}}$). Taking the
CO mole fraction as an observational constraint, we can constrain the deep H$_{2}$O abundance on Jupiter for a plausible range of eddy $K_{zz}$ values. For example, our model solutions are
represented by the solid lines in Fig.~\ref{fig: CO sensitivity}.  The slope of each line is proportional to $K_{zz}$ and hence the vertical convective mixing rate.  The observed upper
tropospheric CO mole fraction of $(1.0\pm0.2)\times10^{-9}$ \citep{bezard2002}, represented by the shaded area, constrains the range of $K_{zz}$ and $E_{\textrm{H}_{2}\textrm{O}}$ values in our
model solutions which are consistent with observations of CO in Jupiter's troposphere.

For our nominal model with $K_{zz}$ = $1\times10^{8}$ cm$^{2}$ s$^{-1}$, our results show that an H$_{2}$O abundance of 2.5 times the solar H$_{2}$O/H$_{2}$ ratio best reproduces the observed CO
abundance.  This enrichment corresponds to H$_{2}$O/H$_{2}=2.4\times10^{-3}$ or $X_{\textrm{H}_{2}\textrm{O}}=2.1\times10^{-3}$ in Jupiter's troposphere. If we further consider a range of
plausible $K_{zz}$ values from $4\times10^{7}$ to $1\times10^{9}$ cm$^{2}$ s$^{-1}$ based on the theories described in Section \ref{ss Numerical Approach}, we find that H$_{2}$O/H$_{2}$ ratios
between 0.52 and 5.20 times the solar ratio remain consistent with CO observations, including uncertainties in the observations themselves.

The above discussion does not consider the potential errors due to uncertainties in reaction kinetics and thermodynamic parameters.  The error bars in Fig.~\ref{fig: CO sensitivity} show our
attempt to quantify the effect of these uncertainties.  As described below (\S\ref{ss: Reaction Pathways}), the CO $\rightarrow$ CH$_4$ conversion is largely controlled by one rate-limiting step
(reaction \ref{R863}), for which the rate coefficient was derived using the rate coefficient of the reverse reaction (\ref{R862}), as reported by \citet{jodkowski1999} from ab initio
calculations. However, no uncertainties were discussed for the calculated rate coefficient $k_{\textrm{\ref{R862}}}$. Based upon a literature search of theory-data comparisons for other
reactions, we estimate that $k_{\textrm{\ref{R862}}}$ is uncertain by a factor of $\sim$3, which dominates over uncertainties in the thermodynamic parameters for the calculation of
$k_{\textrm{\ref{R863}}}$. Including this estimated factor-of-three uncertainty in the reaction kinetics, our model results give a water enrichment of 0.3 to 7.3 times the solar abundance,
corresponding to H$_{2}$O/H$_{2} = (0.29 - 7.0)\times10^{-3}$ and $X_{\textrm{H}_{2}\textrm{O}}=(0.25-6.0)\times10^{-3}$. Thus, the subsolar water abundance (0.51x solar) measured by the
\textit{Galileo} entry probe \citep{wong2004} is plausibly consistent with the observed chemical behavior of carbon monoxide if relatively rapid vertical mixing rates (e.g., corresponding to
$K_{zz}\sim1\times10^{9}$ cm$^{2}$ s$^{-1}$) prevail in Jupiter's troposphere.  For comparison, \citet{bezard2002} derive a water abundance of H$_{2}$O/H$_{2}=(0.34-15.3)\times10^{-4}$ (0.4 to
15.9 times the solar abundance) using a timescale approach rather than a full kinetic-transport model. Refinement of our estimate for the water enrichment may require updated rate-constant
measurements for key reactions in the CO reduction mechanism and improved constraints on the $K_{zz}$ profile in Jupiter's troposphere.

\subsection{Reaction Pathways for CO Destruction}\label{ss: Reaction Pathways}

In the present study we did not select the rate-determining step \textit{a priori}, but rather we input a full reaction list and let the code identify the predominant (fastest) chemical pathway
for net CO$\rightarrow$CH$_{4}$ conversion based upon the relative rates of all reactions included in the model.  As discussed above (\S\ref{ss: Chemical Reactions}), reaction rate coefficients
have been obtained from experimental or theoretical data available in the literature or calculated from the rate coefficient of the reverse reaction at each level in the model via
Eq.~(\ref{eq:reaction reversal}). Empirical rate-constant expressions for the important reactions involving CO reduction in our kinetic scheme are listed in Table \ref{tab: reaction pathway}.

Our model results indicate that carbon monoxide reduction to methane in Jupiter's atmosphere is dominated by the following series of reactions
\begin{align}
\textrm{H} + \textrm{CO} & \xrightarrow{\textrm{M}} \textrm{HCO}\tag{R742}\label{R742}\\
\textrm{H}_{2} + \textrm{HCO} & \rightarrow \textrm{H}_{2}\textrm{CO} + \textrm{H}\tag{R787}\label{R787}\\[-1mm]
\textrm{H} + \textrm{H}_{2}\textrm{CO} & \xrightarrow{\textrm{M}} \textrm{CH}_{3}\textrm{O}\tag{R831}\label{R831}\\
\textrm{H}_{2} + \textrm{CH}_{3}\textrm{O} & \rightarrow \textrm{CH}_{3}\textrm{OH}+\textrm{H}\tag{R863}\label{R863}\\
\textrm{H} + \textrm{CH}_{3}\textrm{OH} & \rightarrow \textrm{CH}_{3} + \textrm{H}_{2}\textrm{O}\tag{R858}\label{R858}\\
\textrm{H}_{2} + \textrm{CH}_{3} & \rightarrow \textrm{CH}_{4} + \textrm{H}\tag{R151}\label{R151}\\[-3mm]
\cline{1-2}\textrm{CO} + 3\, \textrm{H}_{2} & \rightarrow \textrm{CH}_{4}+\textrm{H}_{2}\textrm{O}\tag{net}
\end{align}
where M refers to any third body (i.e., the reaction is termolecular) and the reaction numbers are the assigned reaction numbers in our kinetic model (see Table~\ref{tab: reaction pathway}).
 This kinetic scheme differs considerably from that proposed by \citet{prinn1977} and adopted by subsequent authors, which assumes that the rate-limiting step for converting CO to CH$_{4}$ is the
reaction
\begin{equation}\label{R669}
\textrm{H}_{2}+\textrm{H}_{2}\textrm{CO}\rightarrow \textrm{CH}_{3} + \textrm{OH}.\tag{R669}
\end{equation}
\citet{prinn1977} calculated a rate constant for \ref{R669} using the rate of the reaction CH$_{3}$ + OH \citep{fenimore1969,bowman1974}, assuming that formaldehyde is a major product.  However,
a number of experimental and theoretical kinetic investigations over the past three decades demonstrate that the reaction channel forming H$_{2}$CO is insignificant in the reaction between
methyl and hydroxyl radicals and that the rate estimated by \citet{prinn1977} is incorrect
\citep{barnun1985,dean1987,yung1988,deavillezpereira1997,xia2001,krasnoperov2004,baulch2005,jasper2007}, as has been noted in other studies of CO quenching kinetics
\citep{yung1988,griffith1999,bezard2002,cooper2006}.  Updated second-order temperature-dependent rate coefficients for the reverse reaction OH + CH$_{3}$ $\rightarrow$ H$_{2}$CO + H$_{2}$ (R668)
are now available from \citet{dean1987} and \citet{deavillezpereira1997}, and show that the rate coefficient for \ref{R669} is $\sim$3 orders of magnitude smaller than that estimated by
\citet{prinn1977} at quench-level temperatures ($\sim$1000 K) on Jupiter.  Furthermore, we note that \ref{R669} is included in our model reaction list but that its contribution to CO destruction
is insignificant, consistent with kinetics literature \citep[e.g.,][]{fenimore1969,dean1987,deavillezpereira1997,xia2001,krasnoperov2004,baulch2005,jasper2007}, because faster, alternative
reaction pathways exist for CO$\rightarrow$CH$_{4}$ conversion in Jupiter's troposphere.

Our reaction scheme (in particular reactions \ref{R831} through \ref{R151}) is similar to that proposed by \citet{yung1988} and adopted by \citet{griffith1999} for Gliese 229B and
\citet{bezard2002} for Jupiter.  However, we find that H$_2$ + CH$_3$O $\rightarrow$ CH$_3$OH + H (reaction \ref{R863}) is the slowest reaction in this dominant scheme and, as such, is the
rate-limiting step, whereas \citet{yung1988}, \citet{griffith1999}, and \citet{bezard2002} assumed H + H$_2$CO + M $\rightarrow$ CH$_3$O + M (\ref{R831}) was the rate-limiting reaction for CO
destruction.  We also note that the activation energy for \ref{R863} at high temperatures ($\sim$90 kJ mol$^{-1}$ for 1600 to 3600 K) is similar to that measured by \citet{barnun1975} and
\citet{barnun1985} for CO reduction in high-temperature shocks.  However, we are unable to confirm whether CO$\rightarrow$CH$_{4}$ conversion in the shock-tube experiments involves the same
reaction pathways (\ref{R742} through \ref{R151}) as we find for Jupiter \citep[cf.][]{yung1988}.

There are no experimental data available for \ref{R863}, so we have determined its rate from the rate coefficient of the reverse reaction,
\begin{equation}
\textrm{H} + \textrm{CH}_{3}\textrm{OH}  \rightarrow  \textrm{CH}_{3}\textrm{O} + \textrm{H}_{2},\tag{R862}\label{R862}
\end{equation}
as calculated by \citet{jodkowski1999} using transition-state theory.  Adopting $k_{\textrm{\ref{R862}}}$ as $k_{r}$ in equation (\ref{eq:reaction reversal}) along with thermodynamic parameters
for CH$_{3}$OH from \citet{chen1977}, H from \citet{chase1998}, and CH$_{3}$O from \citet{ruscic2005}, we calculated the reaction rate coefficient for \ref{R863} at each temperature level in our
model and determine an empirical fit of the form
\begin{equation}\label{k R863}
k_{\textrm{\ref{R863}}}=1.77\times10^{-22} T^{3.09} e^{(-3055/T)}.
\end{equation}
We note that \citet[][]{jodkowski1999} also calculated $k_{\textrm{\ref{R863}}}$ from the rate coefficient of the reverse reaction, and estimate $k_{\textrm{\ref{R863}}}=2.10\times10^{-25}
T^{4.0} e^{-2470/T}$.  However, \citet{jodkowski1999} use equilibrium constants derived theoretically from molecular parameters, whereas we have taken advantage of the thermodynamic updates
provided by \citet{ruscic2005}. Our value for $k_{\textrm{\ref{R863}}}$ therefore differs from that of \citet[][]{jodkowski1999}, but by an amount that is much smaller than our overall
factor-of-three estimated uncertainty in $k_{\textrm{\ref{R863}}}$ over the range of temperatures considered in our model.  Our results regarding the quench point, and its constraint on the
water enrichment, are particularly sensitive to the rate coefficient adopted for \ref{R863}, and the factor-of-three uncertainty in $k_{\ref{R863}}$ has been included in the overall
uncertainties in the deep water abundance we derive for Jupiter (see \S3.1 and the error bars in Fig.~\ref{fig: CO sensitivity}).    However, there are several alternative reactions, including
\ref{R831}, which may dominate if this rate is in serious error, in which case we would expect qualitatively similar results for CO quench chemistry in Jupiter's troposphere (see note added in
proof).

\subsection{Validity of the Time-scale Approach}\label{ss: Time Constant}

Having identified the dominant chemical mechanism for CO $\rightarrow$ CH$_{4}$ conversion in the Jovian troposphere, we can now test the validity of the time-scale approach previously used for
estimating Jupiter's deep water inventory \citep[e.g.,][]{prinn1977,fegley1988,bezard2002}. The chemical lifetime for CO is given by the expression
\begin{equation}
t_{chem}\textrm{(CO)}= \frac{[\textrm{CO}]}{-d[\textrm{CO}]/dt} = \frac{\textrm{[CO]}}{k_{\textrm{\ref{R863}}}[\textrm{H}_{2}][\textrm{CH}_{3}\textrm{O}]} ,
\end{equation}
assuming reaction \ref{R863} is the rate-limiting reaction.  The vertical mixing time scale is given by
\begin{equation}
t_{mix}=\frac{L^{2}}{K_{zz}}
\end{equation}
where $K_{zz}$ is the eddy diffusion coefficient and $L$ is the characteristic length scale over which the mixing operates.  The atmospheric pressure scale height is traditionally used for $L$
for these types of calculations. However, \citet{smith1998} demonstrated theoretically that $L \approx H$ is not appropriate and  may lead to over-estimates of the mixing length and the mixing
time scale. Using the procedure recommended by \citet{smith1998}, we obtain $L\sim0.12 H$ for CO quenching kinetics on Jupiter.

We have calculated the CO abundance at the quench level (i.e., where $t_{chem} = t_{mix}$) over a range of water enrichments (0.51x to 8x) and $K_{zz}$ values ($4\times10^{7}$ to $1\times10^{9}$
cm$^{3}$ s$^{-1}$) for comparison with our kinetic model results.  For $K_{zz}=1\times10^{8}$ cm$^{2}$ s$^{-1}$ and $X_{\textrm{CO}}=(1.0\pm0.2)\times10^{-9}$ \citep{bezard2002}, our time-scale
approach yields a water enrichment of $2.9\pm0.6$ times solar, whereas our nominal kinetic-transport model yields $E_{\textrm{H}_{2}\textrm{O}}=2.5\pm0.5$ (not including uncertainties in
$K_{zz}$ and $k_{\textrm{\ref{R863}}}$).  Given the overall additional uncertainties in atmospheric mixing rates and reaction kinetics, we conclude that the back-of-the-envelope time-scale
approach gives a reasonably accurate estimate of the H$_{2}$O abundance on Jupiter, provided that the vertical mixing length scale $L$ advocated by \citet{smith1998} is used and that the
appropriate rate-limiting reaction and rate coefficient are considered. Regarding the latter point, we reemphasize that \ref{R863} is the rate-limiting reaction for CO destruction in our model,
in contrast to the assumptions of previous investigators.

\subsection{Chemistry of Other C--O Gases}

Carbon monoxide is not the only species that will undergo quenching in Jupiter's atmosphere.  In principle, any atmospheric constituent subject to vertical transport and reaction chemistry will
quench if the characteristic time scale for convective mixing becomes shorter than the characteristic time scale for kinetic destruction. Here we examine the chemical behavior of other oxidized
carbon gases in Jupiter's troposphere.  Figure \ref{fig: carboxy gases} shows the vertical abundance profiles for C--O species for a model in which $K_{zz}=1\times10^{8}$ cm$^{2}$ s$^{-1}$ and
$E_{\textrm{H}_{2}\textrm{O}}=1$. The dotted gray lines indicate abundances predicted by thermochemical equilibrium using the NASA CEA code; divergence from equilibrium is evident for CO,
CO$_{2}$, H$_{2}$CO and CH$_{3}$O and illustrates where rapid vertical mixing and slow reaction kinetics (relative to one another) drives each species toward a constant quenched mole-fraction
profile. Again, CH$_{4}$ and H$_{2}$O remain the dominant carbon- and oxygen-bearing gases, respectively, throughout Jupiter's troposphere.  The CO abundance begins to diverge from equilibrium
near the 640-bar level and assumes a constant quenched profile at altitudes above the 200-bar level.

As pointed out by \citet{prinn1977}, reactions among oxidized (CO, CO$_{2}$, etc.)~or reduced (CH$_{4}$,CH$_{3}$, etc.)~carbon-bearing gases are expected to be much faster than reactions between
these two families. As a result, the chemistry of many oxidized carbon gases is strongly tied to the chemical behavior of CO, the most abundant C--O compound in Jupiter's atmosphere.  For
example, CO quenching (via the rate-limiting step \ref{R863}) immediately affects the vertical abundance profile of CO$_{2}$.  At altitudes above the level where CO begins to quench ($\sim$640
bar), CO$_{2}$ remains in equilibrium with CO via the reactions
\begin{align}
\textrm{OH} + \textrm{CO} & \rightarrow \textrm{CO}_{2} + \textrm{H}\tag{R708}\\
\textrm{H} + \textrm{CO}_{2} & \rightarrow \textrm{CO} + \textrm{OH}\tag{R709} \quad .
\end{align}
The CO$_{2}$ abundance thus slightly \emph{increases} with altitude until CO$\rightleftarrows$CO$_{2}$ conversion (via R708$\rightleftarrows$R709) itself quenches near the 500-bar level,
whereupon CO$_{2}$ assumes a constant vertical profile to higher altitudes.  Our upper-tropospheric result for CO$_{2}$ is similar to that of \citet{lellouch2002}, who used a time-scale approach
to derive a quenched CO$_{2}$ mole fraction abundance of $3\times10^{-12}$ for a solar O/H ratio in Jupiter's troposphere.

Although methanol is one of the products of the rate-limiting reaction (\ref{R863}) in our predominant kinetic scheme for CO$\rightarrow$CH$_{4}$ destruction, the vertical abundance profile for
CH$_{3}$OH shows no dependence upon the chemical behavior of CO.  This result can be explained by comparing the most important reactions for CH$_{3}$OH production at the 400-bar level (for
example) in our nominal model:
\begin{align}
\textrm{H}_{2}\textrm{O} + \textrm{CH}_{3} & \rightarrow \textrm{CH}_{3}\textrm{OH} + \textrm{H} & 93.5\% \tag{R859}\\
\textrm{H}_{2} + \textrm{CH}_{2}\textrm{OH} & \rightarrow \textrm{CH}_{3}\textrm{OH} + \textrm{H} & 6.3\% \tag{R861}\\
\textrm{H}_{2} + \textrm{CH}_{3}\textrm{O} & \rightarrow \textrm{CH}_{3}\textrm{OH}+\textrm{H} & 0.3\% \tag{R863}
\end{align}
Methanol production from CH$_{3}$ and H$_{2}$O via R859 far exceeds production via R863, so the chemical behavior of CH$_{3}$OH is largely decoupled from the CO$\rightarrow$CH$_{4}$ kinetic
scheme (i.e., the carbon in CH$_{3}$OH comes predominantly from CH$_{4}$$\rightarrow$CH$_{3}$ rather than CO, and the oxygen comes from H$_{2}$O).  The CH$_{3}$OH abundance follows an
equilibrium profile until production via R859 quenches near the 200-bar level, at which point vertical mixing drives CH$_{3}$OH toward a constant mole-fraction profile.

The vertical abundance profile for H$_{2}$CO is more complex than that for the other C--O gases.  At altitudes above the level where CO begins to quench, HCO remains in equilibrium with CO via
\ref{R742}$\rightleftarrows$R743 (see Table \ref{tab: reaction pathway}).  In turn, the H$_{2}$CO abundance departs from its predicted equilibrium abundance profile and instead remains in
approximate equilibrium with CO because formaldehyde production via the reactions (at the 400-bar level):
\begin{align}
\textrm{H}_{2} + \textrm{HCO} & \rightarrow \textrm{H}_{2}\textrm{CO} + \textrm{H} & 82.5\% \tag{R787}\\
\textrm{CH}_{3}\textrm{O} & \xrightarrow{\textrm{M}} \textrm{H}_{2}\textrm{CO} + \textrm{H} & 17.5\% \tag{R830}
\end{align}
is balanced by destruction via the reactions:
\begin{align}
\textrm{H}_{2}\textrm{CO} + \textrm{H} & \rightarrow \textrm{HCO} + \textrm{H}_{2} & 81.1\% \tag{R786}\\
\textrm{H} + \textrm{H}_{2}\textrm{CO} & \xrightarrow{\textrm{M}} \textrm{CH}_{3}\textrm{O} & 18.9\% \tag{R831}
\end{align}
This balance continues until conversion via R830$\rightleftarrows$R831 begins to quench near the 500 bar level and can no longer offset production and loss via HCO (R787$\rightleftarrows$R786),
whereupon vertical atmospheric mixing drives H$_{2}$CO toward a constant quenched profile near the 100-bar level (a difference in altitude of roughly one pressure scale height).  As a result
(see Fig.~\ref{fig: carboxy gases}), the H$_{2}$CO vertical abundance profile shows changes in response to two separate quenching reactions (R863 and R830) in Jupiter's troposphere.

\section{Discussion}\label{s: Discussion}

\subsection{Implications for Jupiter's Total Oxygen Inventory}

Water vapor is the dominant oxygen-bearing gas throughout Jupiter's atmosphere and is much more abundant than other oxygen-bearing gases (CO, OH, etc.).  For this reason, the H$_{2}$O abundance
in the troposphere is expected to be representative of the majority of Jupiter's total oxygen inventory. However, some oxygen ($\sim$20\%) is removed from the gas phase by oxide formation (rock)
in the deep atmosphere \citep[e.g.,][]{fegley1988,lodders2004apj,visscher2005}.  This removal must be considered before evaluating the bulk planetary oxygen abundance and, in turn, the
heavy-element composition of planetesimals during Jupiter's formation.

The fraction of oxygen removed by rock depends upon the abundance of all rock-forming elements (Mg, Si, Ca, Al, Na, K, Ti) relative to the total oxygen abundance. Following the method of
\citet{visscher2005}, and using updated solar abundances from \citet{lodders2009}, the relative abundances of water vapor, total oxygen, and the rock-forming elements can be written as
\begin{equation}\label{eq: enrichment mass balance}
E_{\textrm{H}_{2}\textrm{O}}=1.261E_{\Sigma{\textrm{O}}}-0.261E_{\textrm{rock}}
\end{equation}
where $E_{i}$ represents the enrichment (over solar ratios) for each component $i$.  This expression serves as a general mass-balance constraint for the relative abundances of water, oxygen, and
rock over a range of heavy element enrichments in Jupiter's interior \citep{visscher2005}.  To derive this expression, it was assumed that all of the rock-forming elements are equally enriched,
and $E_{\textrm{rock}}=2.74\pm0.65$ was adopted based upon the ``deep'' tropospheric abundance of sulfur \citep[as H$_{2}$S;][]{wong2004}, which behaves as a rock-forming element in meteorites
\citep[][]{lodders2004apj}. Using $E_{\textrm{rock}}=2.74\pm0.65$ in equation (\ref{eq: enrichment mass balance}) along with the \textit{Galileo} entry probe H$_{2}$O abundance of
$E_{\textrm{H}_{2}\textrm{O}}=0.51\pm0.17$ (see Table \ref{tab: gas abundances}) yields a total oxygen abundance (characterized by $E_{\Sigma\textrm{O}}$) of $0.97\pm0.40$ times the solar
abundance of $\Sigma\textrm{O}/\textrm{H}_{2}=1.212\times10^{-3}$.

Using our nominal model result of $E_{\textrm{H}_{2}\textrm{O}}=2.5\pm0.5$, equation (\ref{eq: enrichment mass balance}) yields a total oxygen enrichment of $E_{\Sigma\textrm{O}}=2.5\pm0.8$.
Further considering a 3x uncertainty in reaction kinetics and a range of plausible $K_{zz}$ values from $4\times10^{7}$ to $1\times10^{9}$ cm$^{2}$ s$^{-1}$, our water constraint (0.3 -- 7.3x
solar) gives a total oxygen inventory of 0.7 to 6.5 times the solar $\Sigma\textrm{O}/\textrm{H}_{2}$ ratio in Jupiter's interior.  This value represents the bulk oxygen inventory of Jupiter's
interior consistent with CO chemistry, and includes oxygen as water plus oxygen bound in rock.

\subsection{Implications for Planetary Formation}

All viable giant-planet formation models must consider how heavy elements become entrained in the planet during its formation and evolution.  An important constraint for such models is therefore
the observed atmospheric abundances of gases such as CH$_{4}$, NH$_{3}$, H$_{2}$S, PH$_{3}$, and H$_{2}$O, which are taken to represent the planetary elemental inventories of C, N, S, P, and the
majority of planetary oxygen, respectively.  As illustrated in Fig.~\ref{fig: enrichments}, observations of Jupiter's atmosphere show that C, N, S, P, Ar, Kr, and Xe are enhanced relative to
solar element-to-hydrogen ratios \citep{mahaffy2000,wong2004,lodders2004apj} by factors of 2-4. The enrichment in heavy elements is generally believed to be consistent with the core-accretion
model for giant planet formation \citep{mizuno1980}, in which a rock or rock-ice core initially forms and continues to grow through the accretion of solid planetesimals until it is massive
enough to capture nebular gas \citep{bodenheimer1986,lissauer1987,pollack1996}. In this scenario, the observed heavy element enrichments on Jupiter arise from degassing of the initial core
material and the continued accretion of solid planetesimals, which will most likely vaporize before reaching the core \citep[e.g.,][]{pollack1986}.  What remains unclear is the source and
composition of the planetesimals which provided the enrichment, and several scenarios have been proposed to explain the observed heavy-element abundances.

One distinguishing characteristic of Jovian formation scenarios is the predicted water inventory in Jupiter's deep atmosphere.  For example, trapping of heavy elements by hypothetical
solar-composition icy planetesimals \citep{owen1999,atreya2003,owen2006} would be expected to give an enrichment in oxygen (as H$_{2}$O) around 3$\pm$1 times solar, similar that for the other
heavy elements. Trapping of heavy elements in the form of clathrate hydrates near the snow line \citep[e.g.][]{lunine1985,gautier2001,hersant2004} would yield higher water abundances, with
predicted enrichments ranging from $\sim$6 \citep{mousis2009} to $\sim$8 \citep{alibert2005} to as high as $\gtrsim$17 \citep{gautier2001,gautier2001b} or $\sim$19 \citep{hersant2004} times the
solar H$_{2}$O/H$_{2}$ ratio of $9.61\times10^{-4}$. Accretion of carbon-rich planetesimals behind a nebular ``tar line'' \citep{lodders2004apj} would give subsolar water abundances similar to
that observed by the \textit{Galileo} probe (0.51x solar).

Our indirect constraint of a deep water abundance 0.3--7.3 times the solar H$_{2}$O/H$_{2}$ ratio from our kinetic-transport model is plausibly consistent with each of these formation mechanisms
but precludes clathrate-hydrate scenarios that would require large ($>\, $8x) water enrichments in Jupiter's deep atmosphere \citep[e.g.,][]{gautier2001,gautier2001b,hersant2004,alibert2005}.

\section{Summary}\label{s: Summary}
We have developed a comprehensive thermochemical kinetics and diffusion model for Jupiter which correctly transitions between equilibrium chemistry in the deep troposphere and
quenched/disequilibrium chemistry in the upper troposphere.  We use this numerical model to compute the vertical abundance profiles for all carbon- and oxygen-bearing atmospheric constituents
and to explore the chemical behavior of CO and other oxidized carbon species in Jupiter's deep atmosphere. We find that carbon monoxide is reduced to CH$_{4}$ via a mechanism similar to that
proposed by \citet{yung1988}; however, our model indicates that the rate-limiting reaction for CO reduction in Jupiter's atmosphere is H$_2$ + CH$_{3}$O $\rightarrow$ CH$_{3}$OH + H rather than
Yung et al.'s proposed reaction H + CH$_3$O + M $\rightarrow$ CH$_3$OH + M.  We also confirm the original analytic prediction of \citet{prinn1977} that the mole fraction of CO will ``quench''
and remain constant with altitude when kinetic reaction rates can no longer compete with atmospheric mixing.  This quenching occurs at the $\sim$400 bar (1000 K) level in our nominal model.
Carbon monoxide is not the only species to quench; virtually all atmospheric constituents will quench at some point where temperatures become low enough to inhibit the kinetics.

Our kinetic-transport model quantitatively confirms the convenient, back-of-the-envelope time-scale approach currently used to explore quenched disequilibrium chemistry on giant planets and
brown dwarfs \citep[e.g.,][]{prinn1977,lewis1984,fegley1985apj,fegley1988,lodders1994,lodders2002,griffith1999,bezard2002,visscher2005}.  We find that the the time-scale approach is valid for
estimating Jupiter's water inventory, provided that the correct rate-limiting reaction is considered (which we find to be reaction R863, H$_2$ + CH$_{3}$O $\rightarrow$ CH$_{3}$OH + H) and
provided that the mixing length $L$ is determined via the procedure advocated by \citet{smith1998}.

Using the CO abundance reported by \citet{bezard2002} as our observational constraint, our model-data comparisons indirectly constrain the Jovian deep water abundance to lie in the range
0.3--7.3 times the solar H$_{2}$O/H$_{2}$ ratio of $9.61\times10^{-4}$.  Our results suggest that the enrichment for oxygen (as H$_{2}$O) is similar, to within uncertainties, as that for carbon,
nitrogen, and other heavy elements --- giant-planet formation scenarios that require very large ($>\, $8x) enrichments in the water abundance (such as some clathrate-hydrate formation scenarios)
are precluded. The subsolar water abundance (0.51x solar) measured by the \textit{Galileo} entry probe \citep{wong2004} remains plausibly consistent with the observed tropospheric abundance of
carbon monoxide if relatively rapid vertical mixing (e.g., $K_{zz}\ \gtrsim 1\times10^{9}$ cm$^2$ s$^{-1}$)  prevails in Jupiter's deep troposphere.

We will not be able to narrow our estimated range of Jovian deep water enrichments without experimental confirmation of the rate coefficient for the reaction of methoxy with H$_2$ (CH$_3$O +
H$_2$ $\rightarrow$ CH$_3$OH + H) at high pressures, for temperatures near 1000 K, and our results are subject to revision as updated kinetics data become available.  Perhaps more importantly,
we need a better understanding of appropriate diffusion coefficients that can be used to represent convective processes under tropospheric conditions on Jupiter. The \textit{Juno} mission may
provide microwave data \citep{janssen2005,bolton2006} that can be used to test our model prediction regarding the Jovian deep water abundance.

We point out that oxygen species are not the only constituents to quench in our model; the quenching of nitrogen species like N$_{2}$ and HCN is interesting in its own right and will be the
subject of a future investigation. Our kinetics-transport model can easily be applied to other giant planets and brown dwarfs.  Of particular interest is (1) constraining the deep water
abundance on the other giant planets in our own solar system from current or future tropospheric CO observations (which must be able to constrain the vertical profile to separate the
contributions arising from possible internal and external sources), as no current plans to send multiprobe missions to these planets are on schedule for the near future, and (2) predicting the
vertical variation of observable species in brown dwarfs and extrasolar giant planets, as these atmospheres are unlikely to be in complete thermochemical equilibrium.

\section*{Acknowledgements}

This work was supported by the NASA Planetary Atmospheres Program (NNH08ZDA001N) and the Lunar and Planetary Institute/USRA (NASA Cooperative Agreement NCC5-679).  LPI Contribution No.~1546.
\label{lastpage}\\

\textit{{Note added in proof}}-- In our kinetic mechanism for CO $\rightarrow$ CH$_{4}$ conversion in Jupiter's troposphere, we adopted a rate coefficient for the reaction H + CH$_{3}$OH
$\rightarrow$ CH$_{3}$ + H$_{2}$O (R858) based upon the work of \citet{hidaka1989}.  However, recent literature studies suggest that this rate coefficient is inappropriate and may lead to
over-estimates for the rate of (R858).  The three main reaction pathways for H + CH$_{3}$OH are:
\begin{align}
\textrm{H} + \textrm{CH}_{3}\textrm{OH} & \rightarrow \textrm{CH}_{2}\textrm{OH} + \textrm{H}_{2}\tag{R860}\\
\textrm{H} + \textrm{CH}_{3}\textrm{OH} & \rightarrow \textrm{CH}_{3}\textrm{O} + \textrm{H}_{2}\tag{R862}\\
\textrm{H} + \textrm{CH}_{3}\textrm{OH} & \rightarrow \textrm{CH}_{3} + \textrm{H}_{2}\textrm{O}\tag{R858}
\end{align}
Literature values for the relative rates of each pathway are contradictory \citep{lendvay1997,jodkowski1999,baulch2005}. Laboratory and theoretical investigations indicate that (R860) is the
dominant pathway with a rate coefficient that is 4 times \citep{tsang1987,norton1989,norton1990} to $\sim$ 30 times \citep{lendvay1997,jodkowski1999,carvalho2008} greater than that of (R862).
These studies also suggest that (R860) dominates over (R858) by a factor of $\sim 10-40$ at 1000 K \citep[e.g.,][]{aronowitz1977,hoyermann1981,spindler1982,norton1989}, except for
\citet{lendvay1997}, who predict $k_{\textrm{R860}}/k_{\textrm{R858}}\gtrsim 5000$ at 1000 K. Although (R858) is the most exothermic of the three reactions listed above \citep[see
also][]{yung1988}, this pathway appears to be inhibited by a large activation energy \citep[e.g.,][]{lendvay1997}

The rate-limiting step (RLS) for CO $\rightarrow$ CH$_{4}$ (and the chemical behavior of methanol) in Jupiter's troposphere is sensitive to both the overall rate of H + CH$_{3}$OH and the
relative contribution of each reaction pathway.  We tested the sensitivity of our model results to variations in methanol kinetics by adopting a rate coefficient for (R858) using
$k_{\textrm{R860}}/k_{\textrm{R858}}=40$ at 1000 K \citep[e.g.,][]{norton1989}, along with an activation energy of $\sim103$ kJ mol$^{-1}$ \cite[e.g.,][]{lendvay1997}. Depending upon the overall
reaction rate and the relative rates of each pathway, we identify four possible rate-limiting reactions from our model results:
\begin{align}
\textrm{H}_{2} + \textrm{CH}_{3}\textrm{O} & \rightarrow \textrm{CH}_{3}\textrm{OH} + \textrm{H}\tag{R863}\\
\textrm{H} + \textrm{CH}_{3}\textrm{OH} & \rightarrow \textrm{CH}_{3} + \textrm{H}_{2}\textrm{O}\tag{R858}\\[-1mm]
\textrm{H} + \textrm{H}_{2}\textrm{CO} & \xrightarrow{\textrm{M}} \textrm{CH}_{3}\textrm{O}\tag{R831}\\[-1mm]
\textrm{CH}_{3}\textrm{OH} & \xrightarrow{\textrm{M}} \textrm{CH}_{3} + \textrm{OH}\tag{R667}
\end{align}
We note that (R863), (R858), and (R831) are each present in our mechanism for CO destruction in Jupiter's atmosphere, and that the (R831) pathway has been adopted as the rate-limiting step in
previous studies of CO quenching kinetics \citep[][]{yung1988,griffith1999,bezard2002}. The reaction pathway (R667) will become the rate-limiting step only if the rate of (R858) is effectively
negligible \citep[R858 remains the RLS even at relatively high $k_{\textrm{R860}}/k_{\textrm{R858}}$ ratios, e.g.,][]{lendvay1997}.
 Considering a CO mole fraction abundance of $1.0 \pm 0.2\times10^{-9}$ \citep{bezard2002} and a range of $K_{zz}$ values from $4\times10^{7}$ to $1\times10^{9}$ cm$^{2}$ s$^{-1}$, our model
results using different $\textrm{H} + \textrm{CH}_{3}\textrm{OH}$ rate-coefficient data are summarized in Table \ref{tab: noteinproof}. For our preferred model, we take $k_{\textrm{R860}}$ and
$k_{\textrm{R862}}$ from \citet[][]{jodkowski1999} and adopt $k_{\textrm{R858}}\approx9.41\times10^{-9}e^{(-12400/T)}$ cm$^{-3}$ s$^{-1}$ (cf.~Table 3), assuming
$k_{\textrm{R860}}/k_{\textrm{R858}}=40$ at 1000 K. In this approach, (R863) remains the rate-limiting step and the observed CO abundance ($1.0\pm0.2$ ppb) yields a Jovian water abundance of
0.4--3.4 times the solar H$_{2}$O/H$_{2}$ ratio ($9.61\times10^{-4}$), not including uncertainties in reaction kinetics.

In each case, the water abundance derived from CO quench chemistry is consistent with our earlier results (which include an estimated factor-of-three uncertainty in reaction kinetics) of
0.3--7.3 times the solar H$_{2}$O/H$_{2}$ ratio.  The different rate-coefficient data give roughly similar results because other reactions such as (R831) or (R858) may dominate if (R863) is not
the rate-limiting step.  Each of these reactions (R831, R858, R863) quench in the same vicinity (within a fraction of a scale height) in Jupiter's troposphere. Our overall conclusions regarding
the Jovian water inventory, and the reactions given in our kinetic scheme, thus remain unchanged.  We emphasize the need for laboratory measurements of all the pathways for H + CH$_{3}$OH in
order to definitively identify the rate-limiting step for CO $\rightarrow$ CH$_{4}$ in Jupiter's troposphere and to refine estimates of Jupiter's deep water abundance.  We gratefully acknowledge
Greg Smith for alerting us to the uncertainties associated with our adopted rate for the CH$_{3}$ + H$_{2}$O reaction pathway.

\bibliographystyle{elsarticle-harv}

\begin{thebibliography}{}

\bibitem[{{Alibert} et~al.(2005){Alibert}, {Mousis}, and {Benz}}]{alibert2005} {Alibert}, Y., {Mousis}, O., {Benz}, W., 2005.
{On the volatile enrichments and  composition of Jupiter}. \textit{Astrophys. J. Lett.} 622, L145--L148.

\bibitem[{{Allen} et~al.(1981){Allen}, {Yung}, and {Waters}}]{allen1981} {Allen}, M., {Yung}, Y.~L., {Waters}, J.~W., 1981.
{Vertical transport and photochemistry in the terrestrial mesosphere and lower thermosphere (50-120 km)}. \textit{J. Geophys. Res.} 86, 3617--3627.

\bibitem[{{Arai} et~al.(1981){Arai}, {Nagai}, and {Hatada}}]{arai1981} {Arai}, H., {Nagai}, S., {Hatada}, M., 1981.
Radiolysis of methane containing small amounts of carbon monoxide-formation of organic acids. \textit{Radiat. Phys. Chem.} 17, 211--216.

\bibitem[{{Aronowitz} et~al.(1977){Aronowitz}, {Naegeli}, and
  {Glassman}}]{aronowitz1977}
{Aronowitz}, D., {Naegeli}, D.~W., {Glassman}, I., 1977. Kinetics of the
  pyrolysis of methanol. J. Phys. Chem. 81~(25), 2555--2559.

\bibitem[{{Atreya}(2004)}]
{atreya2004} {Atreya}, S.~K., 2004. {Composition, clouds, and origin of Jupiter's atmosphere - a case for deep multiprobes into giant planets}. In: {A.~Wilson} (Ed.), \textit{Planetary Probe
Atmospheric Entry and Descent Trajectory Analysis and Science.} Vol. 544 of ESA Special Publication. pp. 57--62.

\bibitem[{{Atreya} et~al.(2003){Atreya}, {Mahaffy}, {Niemann}, {Wong}, and  {Owen}}]
{atreya2003} {Atreya}, S.~K., {Mahaffy}, P.~R., {Niemann}, H.~B., {Wong}, M.~H., {Owen}, T.~C., 2003. {Composition and origin of the atmosphere of Jupiter - An update, and implications for the
extrasolar giant planets}. \textit{Planet. Space Sci.} 51, 105--112.

\bibitem[{{Bar-Nun} and {Podolak}(1985)}]{barnun1985}
{Bar-Nun}, A., {Podolak}, M., 1985. {The contribution by thunderstorms to
  the abundances of CO, C${_2}$H${_2}$, and HCN on Jupiter}. \textit{Icarus} 64,
  112--124.

  \bibitem[{{Bar-Nun} and {Shaviv}(1975)}]{barnun1975}
{Bar-Nun}, A., {Shaviv}, A., 1975. {Dynamics of the chemical evolution of Earth's primitive atmosphere}. \textit{Icarus} 24,
  197--210.

\bibitem[{{Barshay} and {Lewis}(1978)}]{barshay1978} {Barshay}, S.~S., {Lewis}, J.~S., 1978. {Chemical structure of the deep atmosphere of Jupiter}. \textit{Icarus} 33, 593--611.

\bibitem[{Baulch et~al.(2005)Baulch, Bowman, Cobos, Cox, Just, Kerr, Pilling, Stocker, Troe, Tsang, Walker, and Warnatz}]
{baulch2005} Baulch, D.~L., Bowman, C.~T., Cobos, C.~J., Cox, R.~A., Just, T., Kerr, J.~A., Pilling, M.~J., Stocker, D., Troe, J., Tsang, W., Walker, R.~W., Warnatz, J., 2005. Evaluated kinetic
data for combustion modeling: Supplement ii. \textit{J. Phys. Chem. Ref. Data} 34, 757--1397.

\bibitem[{Baulch et~al.(1994)Baulch, Cobos, Cox, Frank, Hayman, Just, Kerr, Murrells, Pilling, Troe, Walker, and Warnatz}]
{baulch1994} Baulch, D.~L., Cobos, C.~J., Cox, R.~A., Frank, P., Hayman, G., Just, T., Kerr, J.~A., Murrells, T., Pilling, M.~J., Troe, J., Walker, R.~W., Warnatz, J., 1994. Evaluated kinetic
data for combustion modeling. supplement i. \textit{J. Phys. Chem. Ref. Data} 23, 847--848.

\bibitem[{Baulch et~al.(1992)Baulch, Cobos, Esser, Frank, Just, Kerr, Pilling, Troe, Walker, and Warnatz}]
{baulch1992} Baulch, D.~L., Cobos, C. J.~Cox, R.~A., Esser, C., Frank, P., Just, T., Kerr,  J.~A., Pilling, M.~J., Troe, J., Walker, R.~W., Warnatz, J., 1992. Evaluated kinetic data for
combustion modelling. \textit{J. Phys. Chem. Ref. Data} 21, 411--429.

\bibitem[{{Beer}(1975)}]{beer1975}{Beer}, R., Sep. 1975. {Detection of carbon monoxide in Jupiter}. \textit{Astrophys. J. Lett.} 200, L167--L169.

\bibitem[{{Beer} and {Taylor}(1978)}]{beer1978}{Beer}, R., {Taylor}, F.~W., 1978. {The abundance of carbon monoxide in  Jupiter}. \textit{Astrophys. J.} 221, 1100--1109.

\bibitem[{{B{\'e}zard} et~al.(2002){B{\'e}zard}, {Lellouch}, {Strobel},  {Maillard}, and {Drossart}}]
{bezard2002}{B{\'e}zard}, B., {Lellouch}, E., {Strobel}, D., {Maillard}, J.-P., {Drossart},  P., 2002. {Carbon monoxide on Jupiter: Evidence for both internal and external sources}.
\textit{Icarus} 159, 95--111.

\bibitem[{{Bjoraker} et~al.(1986)
{Bjoraker}, {Larson}, and  {Kunde}}]{bjoraker1986}{Bjoraker}, G.~L., {Larson}, H.~P., {Kunde}, V.~G., 1986. {The gas composition  of Jupiter derived from 5 micron airborne spectroscopic
observations}. \textit{Icarus}  66, 579--609.

\bibitem[{{Bodenheimer} and {Pollack}(1986)}]
{bodenheimer1986} {Bodenheimer}, P., {Pollack}, J.~B., 1986. {Calculations of the accretion and evolution of giant planets The effects of solid cores}. \textit{Icarus} 67, 391--408.

\bibitem[{{Bolton} et~al.(2006)}]{bolton2006}{Bolton}, S., {and the Juno Science Team}, 2006. {The Juno New Frontiers Jupiter polar orbiter mission}. \textit{European Planetary Science Congress 2006}, 535.

\bibitem[{{Bowman}(1974)}]{bowman1974}{Bowman}, C.~T. 1974.  {Non-equilibrium radical concentrations in shock-initiated methane oxidation}.
\textit{15th International Symposium on Combustion }(Combustion Institute, Pittsburgh), 869--882.

\bibitem[{{Burcat} and Ruscic(2005)}]{burcat2005}{Burcat}, A., Ruscic, B., 2005.
\textit{Third millenium ideal gas and condensed phase  thermochemical database for combustion with updates from active thermochemical tables.} TAE 960, ANL-05/20, Argonne National Laboratory. 26
pp.

\bibitem[{{Carvalho} et~al.(2008){Carvalho}, {Barauna}, {Machado}, and
  {Roberto-Neto}}]{carvalho2008}
{Carvalho}, E.~F.~V., {Barauna}, A.~N., {Machado}, F.~B.~C., {Roberto-Neto},
  O., 2008. Theoretical calculations of energetics, structures, and rate
  constants for the H + CH$_{3}$OH hydrogen abstraction reactions. Chem. Phys. Lett. 463, 33--37.

\bibitem[{{Chase}(1998)}]{chase1998} {Chase}, M.~W., 1998. {NIST-JANAF} thermochemical tables. \textit{J. Phys. Chem. Ref.  Data}, 28, monograph no. 9. 1951 pp.

\bibitem[{{Chen} et~al.(1977){Chen}, {Wilhoit}, and {Zwolinski}}]
{chen1977}{Chen}, S., {Wilhoit}, R., {Zwolinski}, B., 1977. Thermodynamic properties of  normal and deuterated methanols. \textit{J. Phys. Chem. Ref. Data} 6~(1), 105--112.

\bibitem[{{Cooper} and {Showman}(2006)}]
{cooper2006}{Cooper}, C.~S., {Showman}, A.~P., 2006. {Dynamics and Disequilibrium Carbon Chemistry in Hot Jupiter Atmospheres, with Application to HD 209458b}. \textit{Astrophys. J. Lett.} 649,
1048--1063.

\bibitem[{{da Silva} et~al.(2006){da Silva}, {Bozzelli}, {Sebbar }, and
  {Bockhorn}}]{dasilva2006}
{da Silva}, G., {Bozzelli}, J., {Sebbar }, N., {Bockhorn}, H., 2006.
  {Thermodynamic and Ab Initio Analysis of the Controversial Enthalpy of
  Formation of Formaldehyde}. \textit{ChemPhysChem} 7, 1119--1126.

\bibitem[{{De Avillez Pereira} et~al.(1997){De Avillez Pereira}, {Baulch},
  {Pilling}, {Robertson}, and {Zeng}}]{deavillezpereira1997}
{De Avillez Pereira}, R., {Baulch}, D., {Pilling}, M., {Robertson}, S., {Zeng},
  G., 1997. {Temperature and Pressure Dependence of the Multichannel Rate
  Coefficients for the CH$_{3}$ + OH System}. \textit{J. Phys. Chem. A}
  101, 9681--9693.

\bibitem[{{de Pater} et~al.(2005){de Pater}, {Deboer}, {Marley}, {Freedman},  and {Young}}]
{depater2005}{de Pater}, I., {Deboer}, D., {Marley}, M., {Freedman}, R., {Young}, R., 2005. {Retrieval of water in Jupiter's deep atmosphere using microwave spectra of its brightness
temperature}. \textit{Icarus} 173, 425--438.

\bibitem[{{Dean} and {Westmoreland}(1987)}]{dean1987}{Dean}, A., {Westmoreland}, P., 1987. {Bimolecular QRRK analysis of methyl  radical reactions}. \textit{Int. J. Chem. Kinet.} 19, 207--228.

\bibitem[{{Dean} and {Bozzelli}(2000)}]{dean2000} {Dean}, A.~M., {Bozzelli}, J.~W., 2000. {Combustion Chemistry of Nitrogen}.
In: {Gardiner}, W.~C., J. (Ed.), {\textit{Gas-Phase Combustion Chemistry}}. Springer: New York, pp. 125--342.

\bibitem[{DeMore et~al.(1992)DeMore, Sander, Golden, Hampson, Kurylo, Howard,  Ravishankara, Kolb, and Molina}]
{demore1992}DeMore, W.~B., Sander, S.~P., Golden, D.~M., Hampson, R.~F., Kurylo, M.~J.,  Howard, C.~J., Ravishankara, A.~R., Kolb, C.~J., Molina, M.~J., 1992. {\textit{Chemical Kinetic and
Photochemical Data for Use in Stratospheric Modelling:  Evaluation No. 10}}. JPL Publication 92-20, Jet Propulsion Laboratory,  Pasadena, CA. 269 pp.

\bibitem[{{Drossart} and {Encrenaz}(1982)}]
{drossart1982}{Drossart}, P., {Encrenaz}, T., 1982. {The abundance of water on Jupiter from  the \textit{Voyager} IRIS data at 5 microns}. \textit{Icarus} 52, 483--491.

\bibitem[{{Encrenaz} et~al.(1996){Encrenaz},
{de Graauw}, {Schaeidt},  {Lellouch}, {Feuchtgruber}, {Beintema}, {B\'{e}zard}, {Drossart}, {Griffin},  {Heras}, {Kessler}, {Leech}, {Morris}, {Roelfsema}, {Roos-Serote}, {Salama},
{Vandenbussche}, {Valentijn}, {Davis}, and {Naylor}}]{encrenaz1996}{Encrenaz}, T., {de Graauw}, T., {Schaeidt}, S., {Lellouch}, E.,  {Feuchtgruber}, H., {Beintema}, D.~A., {B\'{e}zard}, B.,
{Drossart}, P.,  {Griffin}, M., {Heras}, A., {Kessler}, M., {Leech}, K., {Morris}, P.,  {Roelfsema}, P.~R., {Roos-Serote}, M., {Salama}, A., {Vandenbussche}, B.,  {Valentijn}, E.~A., {Davis},
G.~R., {Naylor}, D.~A., 1996. {First results of  ISO-SWS observations of Jupiter.} \textit{Astron. Astrophys.} 315, L397--L400.

\bibitem[{{Fegley} and {Prinn}(1983)}]
{fegley1983}{Fegley}, B., Jr., {Prinn}, R.~G., 1983. {Chemical Probes of Saturn's Deep  Atmosphere}. \textit{LPSC Abstracts} 14, 189--190.

\bibitem[{{Fegley} and {Lodders}(1994)}]
{fegley1994}{Fegley}, B., Jr.,  {Lodders}, K., 1994. {Chemical models of the deep atmospheres of Jupiter and Saturn}. \textit{Icarus} 110, 117--154.

\bibitem[{{Fegley} and {Prinn}(1985)}]
{fegley1985apj} {Fegley}, B., Jr.,  {Prinn}, R.~G., 1985. {Equilibrium and nonequilibrium chemistry of Saturn's atmosphere - Implications for the observability of  PH$_{3}$, N$_{2}$, CO, and
GeH$_{4}$}. \textit{Astrophys. J.} 299, 1067--1078.

\bibitem[{{Fegley} and {Prinn}(1988)}]
{fegley1988}{Fegley}, B., Jr.,  {Prinn}, R.~G., 1988. {Chemical constraints on the water and  total oxygen abundances in the deep atmosphere of Jupiter}. \textit{Astrophys. J.} 324, 621--625.

\bibitem[{{Fenimore}(1969)}]
{fenimore1969}{Fenimore}, C., 1969. {Destruction of methane in water gas by reaction of CH$_{3}$ with OH radicals}. \textit{12th International Symposium on Combustion} (Combustion Institute,
Pittsburgh), 463--467.

\bibitem[{Fernando et~al.(1991)Fernando, {Chen}, and {Boyer}}]
{fernando1991} Fernando, H.~J.~S., {Chen}, R.~R., {Boyer}, D.~L., 1991. Effects of rotation on  convective turbulence. \textit{J. Fluid Mech.} 228, 513--547.

\bibitem[{{Flasar} and {Gierasch}(1977)}]
{flasar1977}{Flasar}, F.~M., {Gierasch}, P.~J., 1977. {Eddy diffusivities within Jupiter}. In: {Jones}, A.~V. (Ed.), \textit{Planetary Atmospheres. Proceedings of the  Nineteenth Symposium of
the Royal Society of Canada.} Ottawa: Royal Society of  Canada, p.~85.

\bibitem[{{Flasar} and {Gierasch}(1978)}]
{flasar1978}{Flasar}, F.~M., {Gierasch}, P.~J., 1978. {Turbulent convection within rapidly  rotating superadiabatic fluids with horizontal temperature gradients}.
  \textit{Geophys. Astro. Fluid} 10, 175--212.

\bibitem[{{Gardiner}(2000)}]{gardiner2000}{Gardiner}, W.~C., J. (Ed.), 2000. \textit{Gas-Phase Combustion Chemistry. } Springer-Verlag: New York. 543 pp.

\bibitem[{{Gautier} et~al.(2001{\natexlab{a}}){Gautier}, {Hersant}, {Mousis},  and {Lunine}}]
{gautier2001}{Gautier}, D., {Hersant}, F., {Mousis}, O., {Lunine}, J.~I.,  2001{\natexlab{a}}. {Enrichments in volatiles in Jupiter: A new  interpretation of the \textit{Galileo} measurements}.
\textit{Astrophys. J. Lett.} 550, L227--L230.

\bibitem[{{Gautier} et~al.(2001{\natexlab{b}}){Gautier}, {Hersant}, {Mousis},  and {Lunine}}]
{gautier2001b}{Gautier}, D., {Hersant}, F., {Mousis}, O., {Lunine}, J.~I., Oct.  2001{\natexlab{b}}. {Erratum: Enrichments in Volatiles in Jupiter: A New  Interpretation of the Galileo
Measurements}. \textit{Astrophys. J. Lett.} 559, L183--L183.

\bibitem[{{Gierasch} and {Conrath}(1985)}]
{gierasch1985}{Gierasch}, P.~J., {Conrath}, B.~J., 1985. {Energy conversion processes in the  outer planets}. In: Hunt, G. E. (Ed.), \textit{Recent Advances in Planetary Meteorology.} Cambrdige:
Cambridge Univ. Press, pp. 121--146.

\bibitem[{{Gladstone} et~al.(1996){Gladstone}, {Allen}, and  {Yung}}]
{gladstone1996}{Gladstone}, G.~R., {Allen}, M., {Yung}, Y.~L., 1996. {Hydrocarbon  photochemistry in the upper atmosphere of Jupiter}. \textit{Icarus} 119, 1--52.

\bibitem[{{Gordon} and {McBride}(1994)}]
{gordon1994}{Gordon}, S., {McBride}, B.~J., 1994. Computer program for calculation of  complex chemical equilibrium compositions and applications. NASA Reference Publication 1311. 55 pp.

\bibitem[{{Griffith} and {Yelle}(1999)}]
{griffith1999}{Griffith}, C.~A., {Yelle}, R.~V., 1999. {Disequilibrium chemistry in a brown  dwarf's atmosphere: Carbon monoxide in {G}liese 229B}. \textit{Astrophys. J.  Lett.} 519, L85--L88.

\bibitem[{{Gurvich} et~al.(1989){Gurvich}, {Veyts}, and  {Alcock}}]
{gurvich1989}{Gurvich}, L.~V., {Veyts}, I.~V., {Alcock}, C.~B., 1989. \textit{Thermodynamic Properties of Individual Substances}, 4th Edition, vol. 1, parts 1 and 2. Hemisphere Publishing, New
York. 929 pp.

\bibitem[{{Gurvich} et~al.(1991){Gurvich}, {Veyts}, and  {Alcock}}]
{gurvich1991}{Gurvich}, L.~V., {Veyts}, I.~V., {Alcock}, C.~B., 1991. \textit{Thermodynamic Properties of Individual Substances}, 4th Edition, vol. 2, parts 1 and 2. Hemisphere Publishing, New
York. 952 pp.

\bibitem[{{Gurvich} et~al.(1994){Gurvich}, {Veyts}, and  {Alcock}}]
{gurvich1994}{Gurvich}, L.~V., {Veyts}, I.~V., {Alcock}, C.~B., 1994. \textit{Thermodynamic Properties of Individual Substances}, 4th Edition, vol. 3, parts 1 and 2. CRC Press, Boca Raton, FL.
1136 pp.

\bibitem[{{Hanel} et~al.(1981){Hanel}, {Conrath}, {Herath}, {Kunde}, and  {Pirraglia}}]
{hanel1981}{Hanel}, R., {Conrath}, B., {Herath}, L., {Kunde}, V., {Pirraglia}, J., Sep.  1981. {Albedo, internal heat, and energy balance of Jupiter - Preliminary  results of the Voyager
infrared investigation}. \textit{J. Geophys. Res.} 86, 8705--8712.

\bibitem[{{Hersant} et~al.(2004){Hersant}, {Gautier}, and
  {Lunine}}]{hersant2004}
{Hersant}, F., {Gautier}, D., {Lunine}, J.~I., 2004. {Enrichment in volatiles
  in the giant planets of the solar system}. \textit{Planet. Space Sci.} 52,
  623--641.

\bibitem[{{Hidaka} et~al.(1989){Hidaka}, {Oki}, and {Kawano}}]{hidaka1989}
{Hidaka}, Y., {Oki}, T., {Kawano}, H., 1989. Thermal decomposition of methanol
  in shock waves. \textit{J. Phys. Chem.} 93~(20), 7134--7139.

\bibitem[{{Hoyermann} and {Wagner}(1981)}]{hoyermann1981}
{Hoyermann}, K.~{Sievert}, R., {Wagner}, H.~G., 1981. Mechanism of the reaction
  of H atoms with methanol. Ber. Bunsenges. Phys. Chem. 85, 149--153.

\bibitem[{Irdam et~al.(1993)Irdam, Kiefer, Harding, and Wagner}]{irdam1993}
Irdam, E., Kiefer, J., Harding, L., Wagner, A., 1993. The formaldehyde
  decomposition chain mechanism. \textit{Int. J. Chem. Kinet.} 25, 285 -- 303.

\bibitem[{{Irwin} et~al.(1998){Irwin}, {Weir}, {Smith}, {Taylor}, {Lambert},
  {Calcutt}, {Cameron-Smith}, {Carlson}, {Baines}, {Orton}, {Drossart},
  {Encrenaz}, and {Roos-Serote}}]{irwin1998}
{Irwin}, P.~G.~J., {Weir}, A.~L., {Smith}, S.~E., {Taylor}, F.~W., {Lambert},
  A.~L., {Calcutt}, S.~B., {Cameron-Smith}, P.~J., {Carlson}, R.~W., {Baines},
  K., {Orton}, G.~S., {Drossart}, P., {Encrenaz}, T., {Roos-Serote}, M., 1998.
  {Cloud structure and atmospheric composition of Jupiter retrieved from
  \textit{Galileo} near-infrared mapping spectrometer real-time spectra}.
  \textit{J. Geophys. Res.} 103, 23001--23022.

\bibitem[{{Janssen} et~al.(2005){Janssen}, {Hofstadter}, {Gulkis}, {Ingersoll},
  {Allison}, {Bolton}, {Levin}, and {Kamp}}]{janssen2005}
{Janssen}, M.~A., {Hofstadter}, M.~D., {Gulkis}, S., {Ingersoll}, A.~P.,
  {Allison}, M., {Bolton}, S.~J., {Levin}, S.~M., {Kamp}, L.~W., Feb. 2005.
  {Microwave remote sensing of Jupiter's atmosphere from an orbiting
  spacecraft}. \textit{Icarus} 173, 447--453.

\bibitem[{{Jasper} et~al.(2007){Jasper}, {Klippenstein}, {Harding}, and
  {Ruscic}}]{jasper2007}
{Jasper}, A., {Klippenstein}, S., {Harding}, L., {Ruscic}, B., 2007. {Kinetics
  of the Reaction of Methyl Radical with Hydroxyl Radical and Methanol
  Decomposition}. \textit{J. Phys. Chem. A} 111, 3932--3950.

\bibitem[{Jodkowski et~al.(1999)Jodkowski, Rayez, Rayez, B\'{e}rces, and
  D\'{o}b\'{e}}]{jodkowski1999}
Jodkowski, J., Rayez, M.-T., Rayez, J.-C., B\'{e}rces, T., D\'{o}b\'{e}, S.,
  1999. Theoretical study of the kinetics of the hydrogen abstraction from
  methanol. 3. reaction of methanol with hydrogen atom, methyl, and hydroxyl
  radicals. \textit{J. Phys. Chem. A} 103, 3750--3765.

\bibitem[{Korobeinichev et~al.(2000)Korobeinichev, Ilyin, Bolshova,
  Shvartsberg, and Chernov}]{korobeinichev2000}
Korobeinichev, O.~P., Ilyin, S.~B., Bolshova, T.~A., Shvartsberg, V.~M.,
  Chernov, A.~A., 2000. {The chemistry of the destruction of organophosphorus
  compounds in flames--III: the destruction of DMMP and TMP in a flame of
  hydrogen and oxygen}. \textit{Combust. Flame} 121, 593--609.

\bibitem[{{Krasnoperov} and {Michael}(2004)}]{krasnoperov2004}
{Krasnoperov}, L., {Michael}, J., 2004. {High-Temperature Shock Tube Studies
  Using Multipass Absorption: Rate Constant Results for OH + CH$_{3}$, OH +
  CH$_{2}$, and the Dissociation of CH$_{3}$OH}. \textit{J. Phys. Chem.
  A} 108, 8317--8323.

\bibitem[{{Kunde} et~al.(1982){Kunde}, {Hanel}, {Maguire}, {Gautier},
  {Baluteau}, {Marten}, {Chedin}, {Husson}, and {Scott}}]{kunde1982}
{Kunde}, V., {Hanel}, R., {Maguire}, W., {Gautier}, D., {Baluteau}, J.~P.,
  {Marten}, A., {Chedin}, A., {Husson}, N., {Scott}, N., 1982. {The
  tropospheric gas composition of Jupiter's north equatorial belt (NH$_{3}$,
  PH$_{3}$, CH$_{3}$D, GeH$_{4}$, H$_{2}$O) and the Jovian D/H isotopic ratio}.
  \textit{Astrophys. J.} 263, 443--467.

\bibitem[{{Larson} et~al.(1975){Larson}, {Fink}, {Treffers}, and
  {Gautier}}]{larson1975}
{Larson}, H.~P., {Fink}, U., {Treffers}, R., {Gautier}, T.~N., 1975. {Detection
  of water vapor on Jupiter}. \textit{Astrophys. J. Lett.} 197, L137--L140.

\bibitem[{{Larson} et~al.(1978){Larson}, {Fink}, and {Treffers}}]{larson1978}
{Larson}, H.~P., {Fink}, U., {Treffers}, R.~C., Feb. 1978. {Evidence for CO in
  Jupiter's atmosphere from airborne spectroscopic observations at 5 microns}.
  \textit{Astrophys. J.} 219, 1084--1092.

\bibitem[{{Lellouch} et~al.(1989){Lellouch}, {Drossart}, and
  {Encrenaz}}]{lellouch1989}
{Lellouch}, E., {Drossart}, P., {Encrenaz}, T., 1989. {A new analysis of the
  Jovian 5-micron \textit{Voyager}/IRIS spectra}. \textit{Icarus} 77, 457--465.

\bibitem[{{Lellouch} et~al.(2002){Lellouch}, {B{\'e}zard}, {Moses}, {Davis},
  {Drossart}, {Feuchtgruber}, {Bergin}, {Moreno}, and
  {Encrenaz}}]{lellouch2002}
{Lellouch}, E., {B{\'e}zard}, B., {Moses}, J.~I., {Davis}, G.~R., {Drossart},
  P., {Feuchtgruber}, H., {Bergin}, E.~A., {Moreno}, R., {Encrenaz}, T.,
  2002. {The Origin of Water Vapor and Carbon Dioxide in Jupiter's
  Stratosphere}. \textit{Icarus} 159, 112--131.

\bibitem[{{Lendvay} et~al.(1997){Lendvay}, {B\'{a}rces}, and
  {M\'{a}rta}}]{lendvay1997}
{Lendvay}, G., {B\'{a}rces}, T., {M\'{a}rta}, F., 1997. An ab initio study of
  the three-channel reaction between methanol and hydrogen atoms: BAC-MP4 and
  Gaussian-2 calculations. J. Phys. Chem. A 101, 1588--1594.

\bibitem[{{Lewis} and {Fegley}(1984)}]{lewis1984}
{Lewis}, J.~S., {Fegley}, M.~B., Jr. 1984. {Vertical distribution of
  disequilibrium species in Jupiter's troposphere}. \textit{Space Sci. Rev.} 39,
  163--192.

\bibitem[{{Li} and {Williams}(1996)}]{li1996}
{Li}, S.~C., {Williams}, F.~A., 1996. Experimental and numerical studies of
  two-stage methanol flames. 26th Symposium (International) on Combustion, The
  Combustion Institute, Pittsburgh, pp.~1017--1024.

\bibitem[{{Lissauer}(1987)}]{lissauer1987}
{Lissauer}, J.~J., 1987. {Timescales for planetary accretion and the structure
  of the protoplanetary disk}. \textit{Icarus} 69, 249--265.

\bibitem[{{Lodders}(2003)}]{lodders2003}
{Lodders}, K., 2003. {Solar System abundances and condensation temperatures of
  the elements}. \textit{Astrophys. J.} 591, 1220--1247.

\bibitem[{{Lodders}(2004)}]{lodders2004apj}
{Lodders}, K., 2004. {Jupiter formed with more tar than ice}. \textit{Astrophys. J.} 611, 587--597.

\bibitem[{{Lodders} and {Fegley}(1994)}]{lodders1994}
{Lodders}, K., {Fegley}, B., Jr., 1994. {The origin of carbon monoxide in
  Neptunes's atmosphere}. \textit{Icarus} 112, 368--375.

\bibitem[{{Lodders} and {Fegley}(2002)}]{lodders2002}
{Lodders}, K., {Fegley}, B., Jr., 2002. {Atmospheric chemistry in giant
  planets, brown dwarfs, and low-mass dwarf stars. I. Carbon, nitrogen, and
  oxygen}. \textit{Icarus} 155, 393--424.

\bibitem[{{Lodders} et~al.(2009){Lodders}, {Palme}, and {Gail}}]{lodders2009}
{Lodders}, K., {Palme}, H., {Gail}, H., 2009. {Abundances of the elements in
  the solar system}. {arXiv: 0901.1149}.

\bibitem[{{Lunine} et~al.(2004){Lunine}, {Coradini}, {Gautier}, {Owen}, and
  {Wuchterl}}]{lunine2004}
{Lunine}, J.~I., {Coradini}, A., {Gautier}, D., {Owen}, T.~C., {Wuchterl}, G.,
  2004. {The origin of Jupiter}. In: Bagenal, F., Dowling, T.~E., McKinnon,
  W.~B. (Eds.), Jupiter.~The Planet, Satellites and Magnetosphere. Cambridge: Cambridge Univ. Press, pp. 19--34.

\bibitem[{{Lunine} and {Stevenson}(1985)}]{lunine1985}
{Lunine}, J.~I., {Stevenson}, D.~J., 1985. {Thermodynamics of clathrate hydrate
  at low and high pressures with application to the outer solar system}. \textit{Astrophys. J. Suppl.}
  58, 493--531.

\bibitem[{{Mahaffy} et~al.(2000){Mahaffy}, {Niemann}, {Alpert}, {Atreya},
  {Demick}, {Donahue}, {Harpold}, and {Owen}}]{mahaffy2000}
{Mahaffy}, P.~R., {Niemann}, H.~B., {Alpert}, A., {Atreya}, S.~K., {Demick},
  J., {Donahue}, T.~M., {Harpold}, D.~N., {Owen}, T.~C., 2000. {Noble gas
  abundance and isotope ratios in the atmosphere of Jupiter from the
  \textit{Galileo} Probe Mass Spectrometer}. \textit{J. Geophys. Res.}
  105, 15061--15072.

\bibitem[{{Miller} et~al.(2005){Miller}, {Pilling}, and {Troe}}]{miller2005}
{Miller}, J., {Pilling}, M., {Troe}, J., 2005. Unravelling combustion
  mechanisms through a quantitative understadning of elementary reactions.
  Proc. Combust. Inst. 30, 43--88.

\bibitem[{{Mizuno}(1980)}]{mizuno1980}
{Mizuno}, H., 1980. {Formation of the giant planets}. \textit{Prog. Theor.
  Phys.} 64, 544--557.

\bibitem[{{Moses} et~al.(1995{\natexlab{a}}){Moses}, {Allen}, and
  {Gladstone}}]{moses1995a}
{Moses}, J.~I., {Allen}, M., {Gladstone}, G.~R., 1995{\natexlab{b}}. {Post-SL9
  sulfur photochemistry on Jupiter}. \textit{Geophys.~Res.~Lett.} 22, 1597--1600.

  \bibitem[{{Moses} et~al.(1995{\natexlab{b}}){Moses}, {Allen}, and
  {Gladstone}}]{moses1995b}
{Moses}, J.~I., {Allen}, M., {Gladstone}, G.~R., 1995{\natexlab{a}}. {Nitrogen
  and oxygen photochemistry following SL9}. \textit{Geophys.~Res.~Lett.} 22, 1601--1604.

\bibitem[{{Moses} et~al.(2000{\natexlab{a}}){Moses}, {B{\'e}zard}, {Lellouch},
  {Gladstone}, {Feuchtgruber}, and {Allen}}]{moses2000a}
{Moses}, J.~I., {B{\'e}zard}, B., {Lellouch}, E., {Gladstone}, G.~R.,
  {Feuchtgruber}, H., {Allen}, M., 2000{\natexlab{a}}. {Photochemistry of
  Saturn's atmosphere. I. Hydrocarbon chemistry and comparisons with ISO
  observations}. \textit{Icarus} 143, 244--298.

\bibitem[{{Moses} et~al.(2005){Moses}, {Fouchet}, {B{\'e}zard}, {Gladstone},
  {Lellouch}, and {Feuchtgruber}}]{moses2005}
{Moses}, J.~I., {Fouchet}, T., {B{\'e}zard}, B., {Gladstone}, G.~R.,
  {Lellouch}, E., {Feuchtgruber}, H., 2005. {Photochemistry and diffusion in
  Jupiter's stratosphere: Constraints from ISO observations and comparisons
  with other giant planets}. \textit{J. of Geophys. Res.--Planet} 110,
  E08001.

\bibitem[{{Moses} et~al.(2000{\natexlab{b}}){Moses}, {Lellouch}, {B{\'e}zard},
  {Gladstone}, {Feuchtgruber}, and {Allen}}]{moses2000b}
{Moses}, J.~I., {Lellouch}, E., {B{\'e}zard}, B., {Gladstone}, G.~R.,
  {Feuchtgruber}, H., {Allen}, M., 2000{\natexlab{b}}. {Photochemistry of
  Saturn's atmosphere. II. Effects of an influx of external oxygen}. \textit{Icarus}
  145, 166--202.

\bibitem[{{Mousis} et~al.(2009){Mousis}, {Marboeuf}, {Lunine}, {Alibert},
  {Fletcher}, {Orton}, {Pauzat}, and {Ellinger}}]{mousis2009}
{Mousis}, O., {Marboeuf}, U., {Lunine}, J.~I., {Alibert}, Y., {Fletcher},
  L.~N., {Orton}, G.~S., {Pauzat}, F., {Ellinger}, Y., 2009. {Determination of
  the Minimum Masses of Heavy Elements in the Envelopes of Jupiter and Saturn}.
  \textit{Astrophys. J.} 696, 1348--1354.

\bibitem[{{Niemann} et~al.(1998){Niemann}, {Atreya}, {Carignan}, {Donahue},
  {Haberman}, {Harpold}, {Hartle}, {Hunten}, {Kasprzak}, {Mahaffy}, {Owen}, and
  {Way}}]{niemann1998}
{Niemann}, H.~B., {Atreya}, S.~K., {Carignan}, G.~R., {Donahue}, T.~M.,
  {Haberman}, J.~A., {Harpold}, D.~N., {Hartle}, R.~E., {Hunten}, D.~M.,
  {Kasprzak}, W.~T., {Mahaffy}, P.~R., {Owen}, T.~C., {Way}, S.~H., Sep. 1998.
  {The composition of the Jovian atmosphere as determined by the Galileo probe
  mass spectrometer}. \textit{J. Geophys. Res.} 103, 22831--22846.

\bibitem[{{Noll} et~al.(1997){Noll}, {Gilmore}, {Knacke}, {Womack}, {Griffith},
  and {Orton}}]{noll1997}
{Noll}, K.~S., {Gilmore}, D., {Knacke}, R.~F., {Womack}, M., {Griffith}, C.~A.,
  {Orton}, G., 1997. {Carbon monoxide in Jupiter after comet Shoemaker-Levy 9}.
  \textit{Icarus} 126, 324--335.

\bibitem[{{Noll} et~al.(1988){Noll}, {Knacke}, {Geballe}, and
  {Tokunaga}}]{noll1988}
{Noll}, K.~S., {Knacke}, R.~F., {Geballe}, T.~R., {Tokunaga}, A.~T., 1988. {The
  origin and vertical distribution of carbon monoxide in Jupiter}.
  \textit{Astrophys. J.} 324, 1210--1218.

\bibitem[{{Norton} and {Dryer}(1989)}]{norton1990}
{Norton}, T.~S., {Dryer}, F.~L., 1989. Some new observations on methanol
  oxidation chemistry. Combust. Sci. Technol. 63, 107--129.

\bibitem[{{Norton} and {Dryer}(1990)}]{norton1989}
{Norton}, T.~S., {Dryer}, F.~L., 1990. Toward a comprehensive mechanism for
  methanol pyrolysis. Int. J. Chem. Kinet. 22, 219--241.

\bibitem[{{Orton} et~al.(1996){Orton}, {Ortiz}, {Baines}, {Bjoraker},
  {Carsenty}, {Colas}, {Dayal}, {Deming}, {Drossart}, {Frappa}, {Friedson},
  {Goguen}, {Golisch}, {Griep}, {Hernandez}, {Hoffmann}, {Jennings},
  {Kaminski}, {Kuhn}, {Laques}, {Limaye}, {Lin}, {Lecacheux}, {Martin},
  {McCabe}, {Momary}, {Parker}, {Puetter}, {Ressler}, {Reyes}, {Sada},
  {Spencer}, {Spitale}, {Stewart}, {Varsik}, {Warell}, {Wild},
  {Yanamandra-Fisher}, {Fazio}, {Hora}, and {Deutsch}}]{orton1996}
{Orton}, G., {Ortiz}, J.~L., {Baines}, K., {Bjoraker}, G., {Carsenty}, U.,
  {Colas}, F., {Dayal}, A., {Deming}, D., {Drossart}, P., {Frappa}, E.,
  {Friedson}, J., {Goguen}, J., {Golisch}, W., {Griep}, D., {Hernandez}, C.,
  {Hoffmann}, W., {Jennings}, D., {Kaminski}, C., {Kuhn}, J., {Laques}, P.,
  {Limaye}, S., {Lin}, H., {Lecacheux}, J., {Martin}, T., {McCabe}, G.,
  {Momary}, T., {Parker}, D., {Puetter}, R., {Ressler}, M., {Reyes}, G.,
  {Sada}, P., {Spencer}, J., {Spitale}, J., {Stewart}, S., {Varsik}, J.,
  {Warell}, J., {Wild}, W., {Yanamandra-Fisher}, P., {Fazio}, G., {Hora}, J.,
  {Deutsch}, L., May 1996. {Earth-Based Observations of the Galileo Probe Entry
  Site}. \textit{Science} 272, 839--840.

\bibitem[{{Orton} et~al.(1998){Orton}, {Fisher}, {Baines}, {Stewart},
  {Friedson}, {Ortiz}, {Marinova}, {Ressler}, {Dayal}, {Hoffmann}, {Hora},
  {Hinkley}, {Krishnan}, {Masanovic}, {Tesic}, {Tziolas}, and
  {Parija}}]{orton1998}
{Orton}, G.~S., {Fisher}, B.~M., {Baines}, K.~H., {Stewart}, S.~T., {Friedson},
  A.~J., {Ortiz}, J.~L., {Marinova}, M., {Ressler}, M., {Dayal}, A.,
  {Hoffmann}, W., {Hora}, J., {Hinkley}, S., {Krishnan}, V., {Masanovic}, M.,
  {Tesic}, J., {Tziolas}, A., {Parija}, K.~C., 1998. {Characteristics of the
  Galileo probe entry site from Earth-based remote sensing observations}. \textit{J. Geophys. Res.}
  103, 22791--22814.

\bibitem[{{Owen} and {Encrenaz}(2006)}]{owen2006}
{Owen}, T., {Encrenaz}, T., 2006. {Compositional constraints on giant planet
  formation}. \textit{Planet. Space Sci.} 54, 1188--1196.

\bibitem[{{Owen} et~al.(1999){Owen}, {Mahaffy}, {Niemann}, {Atreya}, {Donahue},
  {Bar-Nun}, and {de Pater}}]{owen1999}
{Owen}, T., {Mahaffy}, P., {Niemann}, H.~B., {Atreya}, S., {Donahue}, T.,
  {Bar-Nun}, A., {de Pater}, I., 1999. {A low-temperature origin for the
  planetesimals that formed Jupiter}. \textit{Nature} 402, 269--270.

\bibitem[{Page et~al.(1989)Page, Lin, He, and Choudhury}]{page1989}
Page, M., Lin, M.~C., He, Y., Choudhury, T.~K., 1989. Kinetics of the methoxy
  radical decomposition reaction: theory and experiment. \textit{J. Phys. Chem.}
  93~(11), 4404--4408.

\bibitem[{{Pearl} and {Conrath}(1991)}]{pearl1991}
{Pearl}, J.~C., {Conrath}, B.~J., Oct. 1991. {The albedo, effective
  temperature, and energy balance of Neptune, as determined from Voyager data}.
  \textit{J. Geophys. Res.} 96, 18921--18930.

\bibitem[{{Pollack} et~al.(1996){Pollack}, {Hubickyj}, {Bodenheimer},
  {Lissauer}, {Podolak}, and {Greenzweig}}]{pollack1996}
{Pollack}, J.~B., {Hubickyj}, O., {Bodenheimer}, P., {Lissauer}, J.~J.,
  {Podolak}, M., {Greenzweig}, Y., 1996. {Formation of the giant planets by
  concurrent accretion of solids and gas}. \textit{Icarus} 124, 62--85.

\bibitem[{{Pollack} et~al.(1986){Pollack}, {Podolak}, {Bodenheimer}, and
  {Christofferson}}]{pollack1986}
{Pollack}, J.~B., {Podolak}, M., {Bodenheimer}, P., {Christofferson}, B., 1986.
  {Planetesimal dissolution in the envelopes of the forming, giant planets}.
  \textit{Icarus} 67, 409--443.

\bibitem[{{Prinn} and {Barshay}(1977)}]{prinn1977}
{Prinn}, R.~G., {Barshay}, S.~S., 1977. {Carbon monoxide on Jupiter and
  implications for atmospheric convection}. \textit{Science} 198, 1031--1034.

\bibitem[{{Ragent} et~al.(1998){Ragent}, {Colburn}, {Rages}, {Knight}, {Avrin},
  {Orton}, {Yanamandra-Fisher}, and {Grams}}]{ragent1998}
{Ragent}, B., {Colburn}, D.~S., {Rages}, K.~A., {Knight}, T.~C.~D., {Avrin},
  P., {Orton}, G.~S., {Yanamandra-Fisher}, P.~A., {Grams}, G.~W., Sep. 1998.
  {The clouds of Jupiter: Results of the Galileo Jupiter mission probe
  nephelometer experiment}.  \textit{J. Geophys. Res.} 103, 22891--22910.

\bibitem[{{Roos-Serote} et~al.(2004){Roos-Serote}, {Atreya}, {Wong}, and
  {Drossart}}]{roosserote2004}
{Roos-Serote}, M., {Atreya}, S.~K., {Wong}, M.~K., {Drossart}, P., 2004. {On
  the water abundance in the atmosphere of Jupiter}. \textit{Planet. Space
  Sci.} 52, 397--414.

\bibitem[{{Roos-Serote} et~al.(1999){Roos-Serote}, {Drossart}, {Encrenaz},
  {Carlson}, and {Leader}}]{roosserote1999}
{Roos-Serote}, M., {Drossart}, P., {Encrenaz}, T., {Carlson}, R.~W., {Leader},
  F., 1999. {Constraints on the tropospheric cloud structure of Jupiter from
  spectroscopy in the 5-{$\mu$}m region: A Comparison between
  \textit{Voyager}/IRIS, \textit{Galileo}-NIMS, and ISO-SWS Spectra}. \textit{Icarus}
  137, 315--340.

\bibitem[{{Roos-Serote} et~al.(1998){Roos-Serote}, {Drossart}, {Encrenaz},
  {Lellouch}, {Carlson}, {Baines}, {Kamp}, {Mehlman}, {Orton}, {Calcutt},
  {Irwin}, {Taylor}, and {Weir}}]{roosserote1998}
{Roos-Serote}, M., {Drossart}, P., {Encrenaz}, T., {Lellouch}, E., {Carlson},
  R.~W., {Baines}, K.~H., {Kamp}, L., {Mehlman}, R., {Orton}, G.~S., {Calcutt},
  S., {Irwin}, P., {Taylor}, F., {Weir}, A., 1998. {Analysis of Jupiter North
  Equatorial Belt hot spots in the 4-5-{$\mu$}m range from
  \textit{Galileo}/near-infrared mapping spectrometer observations:
  Measurements of cloud opacity, water, and ammonia}.  \textit{J. Geophys. Res.} 103, 23023--23042.

\bibitem[{Ruscic et~al.(2005)Ruscic, Boggs, Burcat, Cs\'{a}sz\'{a}r, Demaison,
  Janoschek, Martin, Morton, Rossi, Stanton, Szalay, Westmoreland, Zabel, and
  B\'{e}rces}]{ruscic2005}
Ruscic, B., Boggs, J.~E., Burcat, A., Cs\'{a}sz\'{a}r, A.~G., Demaison, J.,
  Janoschek, R., Martin, J. M.~L., Morton, M.~L., Rossi, M.~J., Stanton, J.~F.,
  Szalay, P.~G., Westmoreland, P.~R., Zabel, F., B\'{e}rces, T., 2005. {IUPAC
  Critical evalution of thermochemical properties of selected radicals. Part
  I}. \textit{J. Phys. Chem. Ref. Data} 34, 573--656.

\bibitem[{Ruscic et~al.(2002)Ruscic, Wagner, Harding, Asher, Feller, Dixon,
  Peterson, Song, Qian, Ng, Liu, Chen, and Schwenke}]{ruscic2002}
Ruscic, B., Wagner, A.~F., Harding, L.~B., Asher, R.~L., Feller, D., Dixon,
  D.~A., Peterson, K.~A., Song, Y., Qian, X., Ng, C.-Y., Liu, J., Chen, W.,
  Schwenke, D.~W., 2002. On the enthalpy of formation of hydroxyl radical and
  gas-phase bond dissociation energies of water and hydroxyl. \textit{J. Phys. Chem. A} 106, 2727--2747.

\bibitem[{{Seiff} et~al.(1998){Seiff}, {Kirk}, {Knight}, {Young}, {Mihalov},
  {Young}, {Milos}, {Schubert}, {Blanchard}, and {Atkinson}}]{seiff1998}
{Seiff}, A., {Kirk}, D.~B., {Knight}, T.~C.~D., {Young}, R.~E., {Mihalov},
  J.~D., {Young}, L.~A., {Milos}, F.~S., {Schubert}, G., {Blanchard}, R.~C.,
  {Atkinson}, D., 1998. {Thermal structure of Jupiter's atmosphere near the
  edge of a 5-{$\mu$}m hot spot in the north equatorial belt}.  \textit{J. Geophys. Res.} 103, 22857--22890.

\bibitem[{{Smith} et~al.(1999){Smith}, {Golden}, {Frenklach}, {Moriarty},
  {Eiteneer}, {Goldenberg}, {Bowman}, {Hanson}, {Song}, {Gardiner},
  {Lissianski}, and {Qin}}]{smith1999}
{Smith}, G., {Golden}, D., {Frenklach}, M., {Moriarty}, N.~W., {Eiteneer}, B.,
  {Goldenberg}, M., {Bowman}, C., {Hanson}, R., {Song}, S., {Gardiner}, W.,
  {Lissianski}, V., {Qin}, Z., 1999. {GRI-Mech 3.0}. \verb"http://www.me.berkeley.edu/gri-mech/"

\bibitem[{{Smith}(1998)}]{smith1998}
{Smith}, M.~D., 1998. {Estimation of a length scale to use with the quench
  level approximation for obtaining chemical abundances}. \textit{Icarus} 132, 176--184.

\bibitem[{{Spindler} and {Wagner}(1982)}]{spindler1982}
{Spindler}, K., {Wagner}, H.~G., 1982. Zum thermischen unimolekularen zerfall
  von methanol. Ber. Bunsenges. Phys. Chem. 86, 2--13.

\bibitem[{{Sromovsky} et~al.(1998){Sromovsky}, {Collard}, {Fry}, {Orton},
  {Lemmon}, {Tomasko}, and {Freedman}}]{sromovsky1998}
{Sromovsky}, L.~A., {Collard}, A.~D., {Fry}, P.~M., {Orton}, G.~S., {Lemmon},
  M.~T., {Tomasko}, M.~G., {Freedman}, R.~S., Sep. 1998. {Galileo probe
  measurements of thermal and solar radiation fluxes in the Jovian atmosphere}.
   \textit{J. Geophys. Res.} 103, 22929--22978.

\bibitem[{{Stevenson}(1979)}]{stevenson1979}
{Stevenson}, D.~J., 1979. {Turbulent thermal convection in the presence of
  rotation and a magnetic field - A heuristic theory}. \textit{Geophys. and
  Astro. Fluid} 12, 139--169.

\bibitem[{Stone(1976)}]{stone1976}
Stone, P.~H., 1976. The meteorology of the {J}ovian atmosphere. In: Gehrels, T.
  (Ed.), \textit{Jupiter.} Tuscon: Univ. of Arizona Press, pp. 586--618.

\bibitem[{{Taylor} et~al.(2004){Taylor}, {Atreya}, {Encrenaz}, {Hunten},
  {Irwin}, and {Owen}}]{taylor2004}
{Taylor}, F.~W., {Atreya}, S.~K., {Encrenaz}, T., {Hunten}, D.~M., {Irwin},
  P.~G.~J., {Owen}, T.~C., 2004. The composition of the atmosphere of
  {J}upiter. In: Bagenal, F., Dowling, T.~E., McKinnon, W.~B. (Eds.),
  \textit{Jupiter.~The Planet, Satellites and Magnetosphere.} Cambridge: Cambridge
  Univ.~Press, pp. 59--78.

\bibitem[{{Tsang}(1987)}]{tsang1987}
{Tsang}, W., 1987. Chemical kinetic data base for combustion chemistry. part 2.
  methanol. J. Phys. Chem. Ref. Data 16, 471--508.

\bibitem[{{Visscher} and {Fegley}(2005)}]{visscher2005}
{Visscher}, C., {Fegley}, B., Jr., 2005. {Chemical constraints on the water and
  total oxygen abundances in the deep atmosphere of Saturn}. \textit{Astrophys.
  J.} 623, 1221--1227.

\bibitem[{{von Zahn} et~al.(1998){von Zahn}, {Hunten}, and
  {Lehmacher}}]{vonzahn1998}
{von Zahn}, U., {Hunten}, D.~M., {Lehmacher}, G., 1998. {Helium in Jupiter's
  atmosphere: Results from the \textit{Galileo} probe helium interferometer
  experiment}.  \textit{J. Geophys. Res.} 103, 22815--22830.

\bibitem[{{Wong} et~al.(2004){Wong}, {Mahaffy}, {Atreya}, {Niemann}, and
  {Owen}}]{wong2004}
{Wong}, M.~H., {Mahaffy}, P.~R., {Atreya}, S.~K., {Niemann}, H.~B., {Owen},
  T.~C., 2004. {Updated \textit{Galileo} probe mass spectrometer measurements
  of carbon, oxygen, nitrogen, and sulfur on Jupiter}. \textit{Icarus} 171, 153--170.

\bibitem[{{Xia} et~al.(2001){Xia}, {Zhu}, {Lin}, and {Mebel}}]{xia2001}
{Xia}, W.~S., {Zhu}, R.~S., {Lin}, M.~C., {Mebel}, A.~M., 2001. {Low-energy paths for unimolecular decomposition of CH$_{3}$OH: A G2M/statistical theory study}. \textit{Faraday Discussions} 119,
191--205.

\bibitem[{{Yung} et~al.(1988){Yung}, {Drew}, {Pinto}, and {Friedl}}]{yung1988}
{Yung}, Y.~L., {Drew}, W.~A., {Pinto}, J.~P., {Friedl}, R.~R., 1988.
  {Estimation of the reaction rate for the formation of CH$_{3}$O from H +
  H$_{2}$CO - Implications for chemistry in the solar system}. \textit{Icarus} 73,
  516--526.

\end{thebibliography}

\clearpage

\begin{table}[t]
\caption{Gas Abundances in Jupiter's Troposphere}
\small
\begin{tabular*}{1.0\textwidth}{@{\extracolsep{\fill}} ccccc}
\hline
Gas $i$ & Jupiter $i$/H$_{2}$ & Reference & Protosolar $i$/H$_{2}$$^{\textrm{a}}$ & Enrichment Factor\\
\hline
H$_{2}$ & $\equiv1$ & & $\equiv1$ & $\equiv1$\\
He & $0.1574\pm0.0036$ & b, c & 0.1938 & $0.81\pm0.02$\\
CH$_{4}$ & $(2.37\pm0.57)\times10^{-3}$ & d & $5.55\times10^{-4}$ & $4.27\pm1.03$\\
NH$_{3}$ & $(6.64\pm2.54)\times10^{-4}$ & d & $1.64\times10^{-4}$ & $4.05\pm1.55$\\
H$_{2}$O & $(4.9\pm1.6)\times10^{-4}$ & d & $9.61\times10^{-4}$ & $0.51\pm0.17$\\
PH$_{3}$ & $(0.8\pm0.5)\times10^{-6}$ & e & $6.41\times10^{-7}$ & $1.25\pm0.78$\\
H$_{2}$S & $(8.9\pm2.1)\times10^{-5}$ & d & $3.25\times10^{-5}$ & $2.74\pm0.65$\\
Ar & $(1.82\pm0.36)\times10^{-5}$ & f & $7.16\times10^{-6}$ & $2.54\pm0.50$\\
Kr & $(9.30\pm1.70)\times10^{-9}$ & f & $4.31\times10^{-9}$ & $2.16\pm0.39$\\
Xe & $(8.90\pm1.70)\times10^{-10}$ & f & $4.21\times10^{-10}$ & $2.11\pm0.40$\\
\hline  
\end{tabular*}\label{tab: gas abundances}
{\textit{Note}. References: (a) Protosolar abundances from \citet{lodders2009} (b) \citet{niemann1998}, (c) \citet{vonzahn1998}, (d) \citet{wong2004}, (e) 5-$\mu$m spectroscopy, see text (f)
\citet{mahaffy2000}.}
\end{table}

\clearpage

\begin{table}
\caption{Quenched CO Mole Fraction ($X_{\textrm{CO}}$) for Variations in $K_{zz}$ and $E_{\textrm{H}_{2}\textrm{O}}$}
\footnotesize
\begin{tabular*}{1.0\textwidth}{@{\extracolsep{\fill}}ccccc}
\hline H$_{2}$O & \multicolumn{4}{c}{Eddy Diffusion Coefficient $K_{zz}$ (cm$^{2}$ s$^{-1}$)}\\
enrichment$^{a}$ & $4\times10^{7}$ & $1\times10^{8}$ & $4\times10^{8}$ & $1\times10^{9}$\\
\hline
0.51x & $1.16\times10^{-10}$ & $2.02\times10^{-10}$ & $4.57\times10^{-10}$ & $7.71\times10^{-10}$ \\
1x & $2.28\times10^{-10}$ & $3.96\times10^{-10}$ & $8.95\times10^{-10}$ & $1.51\times10^{-9}$ \\
2x & $4.57\times10^{-10}$ & $7.93\times10^{-10}$ & $1.79\times10^{-9}$ & $3.03\times10^{-9}$ \\
4x & $9.18\times10^{-10}$ & $1.59\times10^{-9}$ & $3.60\times10^{-9}$ & $6.08\times10^{-9}$ \\
8x & $1.85\times10^{-9}$ & $3.22\times10^{-9}$ & $7.27\times10^{-9}$ & $1.23\times10^{-8}$ \\
\hline  
\end{tabular*}\label{tab: model results}
\\$^{a}$Water enrichment factor ($E_{\textrm{H}_{2}\textrm{O}}$) relative to a solar H$_{2}$O/H$_{2}$ ratio of
$9.61\times10^{-4}$.  The \textit{Galileo} entry probe measured a deep Jovian water abundance of 0.51x solar \citep{wong2004}.  The observed mole fraction of CO in Jupiter's atmosphere is
$X_{\textrm{CO}}=(1.0\pm0.2)\times10^{-9}$ \citep{bezard2002}.
\end{table}

\clearpage

\begin{table}
\caption{Important Reaction Pathways for CO$\rightleftarrows$CH$_{4}$ in Jupiter's Atmosphere}
\scriptsize
\begin{tabular*}{1.0\textwidth}{@{\extracolsep{\fill}}lr@{}c@{}lll}
\hline
& \multicolumn{3}{c}{Reaction} & Rate Constant & Reference\\
\hline
\textbf{R742} & $\textrm{H} + \textrm{CO}$ & $\xrightarrow{\textrm{M}}$ & $\textrm{HCO}$ & $k_{0}=5.30\times10^{-34}e^{(-370/T)}$ & \citet{baulch1994}\\
     &                            &                            &                & $k_{\infty}=1.96\times10^{-13}e^{(-1366/T})$ & \citet{arai1981}\\
R743 & $\textrm{HCO}$ & $\xrightarrow{\textrm{M}}$ & $\textrm{CO} + \textrm{H}$ & $k_{0}=4.18\times10^{-9}T^{-0.36}e^{(-8294/T)}$ & reverse of R742\\
     &                &                            &                            & $k_{\infty}=1.54\times10^{12}T^{-0.36}e^{(-9290/T})$ & reverse of R742\\
R786 & $\textrm{H} + \textrm{H}_{2}\textrm{CO}$ & $\rightarrow$ & $\textrm{HCO} + \textrm{H}_{2}$ & $k_{\textrm{R786}}=9.53\times10^{-17}T^{1.90}e^{(-1379/T)}$ & \citet{irdam1993}\\
\textbf{R787} & $\textrm{H}_{2} + \textrm{HCO}$ & $\rightarrow$ & $\textrm{H}_{2}\textrm{CO} + \textrm{H}$ & $k_{\textrm{R787}}=3.92\times10^{-19}T^{2.23}e^{(-9082/T)}$ & reverse of R786\\
R830 & $\textrm{CH}_{3}\textrm{O}$ & $\xrightarrow{\textrm{M}}$ & $\textrm{H}_{2}\textrm{CO} + \textrm{H}$ & $k_{0}=1.40\times10^{-6}T^{-1.20}e^{(-7800/T)}$ & \citet{page1989}\\
     &                             &                            &                                          & $k_{\infty}=1.50\times10^{11}Te^{(-12880/T)}$ & \citet{bezard2002}\\
\textbf{R831} & $\textrm{H} + \textrm{H}_{2}\textrm{CO}$ & $\xrightarrow{\textrm{M}}$ & $\textrm{CH}_{3}\textrm{O}$ & $k_{0}=1.28\times10^{-33}T^{-0.30}e^{(3074/T)}$  & reverse of R830 \\
     &                            &                            &                & $k_{\infty}=1.37\times10^{-16}T^{1.90}e^{(-2006/T)}$ & reverse of R830 \\
R862 & $\textrm{H} + \textrm{CH}_{3}\textrm{OH} $ & $\rightarrow$ & $\textrm{CH}_{3}\textrm{O} + \textrm{H}_{2}$ & $k_{\textrm{R862}}=1.135\times10^{-22}T^{3.40}e^{(-3640/T)}$ & \citet{jodkowski1999}\\
\textbf{R863} & $\textrm{H}_{2} + \textrm{CH}_{3}\textrm{O}$ & $\rightarrow$ & $\textrm{CH}_{3}\textrm{OH}+\textrm{H}$  & $k_{\textrm{R863}}=1.77\times10^{-22} T^{3.09} e^{(-3055/T)}$ & reverse of R862\\
\textbf{R858}$^{a}$ & $\textrm{H} + \textrm{CH}_{3}\textrm{OH}$ & $\rightarrow$ & $\textrm{CH}_{3} + \textrm{H}_{2}\textrm{O}$   & $k_{\textrm{R858}}=3.321\times10^{-10}e^{(-2670/T)}$ & \citet{hidaka1989} \\
R859          & $\textrm{H}_{2}\textrm{O} + \textrm{CH}_{3}$ & $\rightarrow$ & $\textrm{CH}_{3}\textrm{OH} + \textrm{H} $   & $k_{\textrm{R859}}=5.83\times10^{-15}Te^{(-15474/T)}$ & reverse of R858\\
R150 & $\textrm{H} + \textrm{CH}_{4}$ & $\rightarrow$ & $\textrm{CH}_{3} + \textrm{H}_{2}$ & $k_{\textrm{R150}}=2.20\times10^{-20}T^{3.00}e^{(-4045/T)}$ & \citet{baulch1992}\\
\textbf{R151} & $\textrm{H}_{2} + \textrm{CH}_{3}$ & $\rightarrow$ & $\textrm{CH}_{4} + \textrm{H}$  & $k_{\textrm{R151}}=8.62\times10^{-24}T^{3.57}e^{(-2995/T)}$ & reverse of R150\\
\hline  
\end{tabular*}\label{tab: reaction pathway}
\\ \textit{Note} -- Reaction numbers in bold refer to steps in the dominant CO$\rightarrow$CH$_{4}$ kinetic scheme. Rate constants for bimolecular reactions ($k_{i}$ for reaction $i$) and
high-pressure limiting rate constants for termolecular reactions ($k_{\infty}$) are in units of cm$^{3}$ s$^{-1}$. Low-pressure limiting rate constants for termolecular reactions ($k_{0}$) are
in units of cm$^{6}$ s$^{-1}$.  For decomposition reactions (e.g., R743 and R830), $k_{0}$ is in units of cm$^{3}$ s$^{-1}$ and $k_{\infty}$ is in units of s$^{-1}$.  Note that although the rate
coefficients were internally reversed within our code, we provide empirical fits to the reaction rate coefficient expressions (valid for $T=$ 400 to 2500 K) for reverse reactions as an aid to
other investigators. $^{a}$See note added in proof.
\end{table}

\clearpage

\begin{table}\label{tab: noteinproof}\label{lasttable}
\caption{Model results for variations in $\textrm{H} + \textrm{CH}_{3}\textrm{OH}$ reaction kinetics} \scriptsize
\begin{tabular*}{1.0\textwidth}{@{\extracolsep{\fill}}ccccc}
\hline
 \multirow{2}{*}{$E_{\textrm{H}_{2}\textrm{O}}$} & \multirow{2}{*}{RLS} & \multicolumn{3}{c}{Rate Coefficients for H + CH$_{3}$OH Reaction Pathways$^{a}$}\\
 & & $k_{\textrm{860}}$ & $k_{\textrm{862}}$ & $k_{\textrm{858}}$\\
\hline
0.4--3.4 & R863 & $2.249\times10^{-21}T^{3.2}e^{(-1755/T)}$ & $1.135\times10^{-22}T^{3.4}e^{(-3640/T)}$ & $9.41\times10^{-9}e^{(-12400/T)}$\\
 & & \citet{jodkowski1999} & \citet{jodkowski1999} & $k_{\textrm{R860}}/k_{\textrm{R858}}=40$ at 1000 K\\
0.5--4.8 & R863 & $5.47\times10^{-15}T^{1.24}e^{(-2260/T)}$ & $2.28\times10^{-16}T^{1.24}e^{(-2260/T)}$ & $1.82\times10^{-8}e^{(-12400/T)}$\\
 & & \multicolumn{2}{c}{\citet{baulch2005}, assuming $k_{\textrm{R860}}/k_{\textrm{R862}}=24^{b}$} & $k_{\textrm{R860}}/k_{\textrm{R858}}=40$ at 1000 K\\
0.6--5.3 &  R858 & $2.72\times10^{-17}T^{2}e^{(-2273/T)}$ & $6.33\times10^{-17}T^{2}e^{(-2947/T)}$ & $1.70\times10^{-8}e^{(-12400/T)}$\\
 & & \multicolumn{2}{c}{\citet{li1996}, $k_{\textrm{R860}}/k_{\textrm{R862}}=0.43e^{674/T}$} & $k_{\textrm{R860}}/k_{\textrm{R858}}=40$ at 1000 K\\
0.1--1.2 & R858 & $4.23\times10^{-10}e^{(-3516/T)}$ & $1.19\times10^{-10}e^{(-5549/T)}$ & $5.08\times10^{-10}e^{(-12400/T)}$\\
 & & \citet{lendvay1997} & \citet{lendvay1997} & \citet{lendvay1997}\\
0.6--5.0 & R858 & $4.56\times10^{-15}T^{1.24}e^{(-2260/T)}$ & $1.14\times10^{-15}T^{1.24}e^{(-2260/T)}$ & $1.52\times10^{-8}e^{(-12400/T)}$\\
 & & \multicolumn{2}{c}{\citet{baulch2005}, assuming $k_{\textrm{R860}}/k_{\textrm{R862}}=4^{c}$} & $k_{\textrm{R860}}/k_{\textrm{R858}}=40$ at 1000 K\\
0.7--6.2 & R831 & $2.82\times10^{-17}T^{2.1}e^{(-2450/T)}$ & $7.04\times10^{-18}T^{2.1}e^{(-2450/T)}$ & $2.94\times10^{-8}e^{(-12400/T)}$\\
 & & \multicolumn{2}{c}{\citet{tsang1987}, $k_{\textrm{R860}}/k_{\textrm{R862}}=4$} & $k_{\textrm{R860}}/k_{\textrm{R858}}=40$ at 1000 K\\
\hline
\end{tabular*}
\\ \textit{Note} -- $^{a}$Rate coefficients are in units of cm$^{3}$ s$^{-1}$ for bimolecular reactions. $^{b}$Based upon \citet{carvalho2008}. $^{c}$Based upon \citet{tsang1987}.
\end{table}

\clearpage

\begin{figure}
\begin{center}
{\includegraphics[angle=0,width=1.0\textwidth]{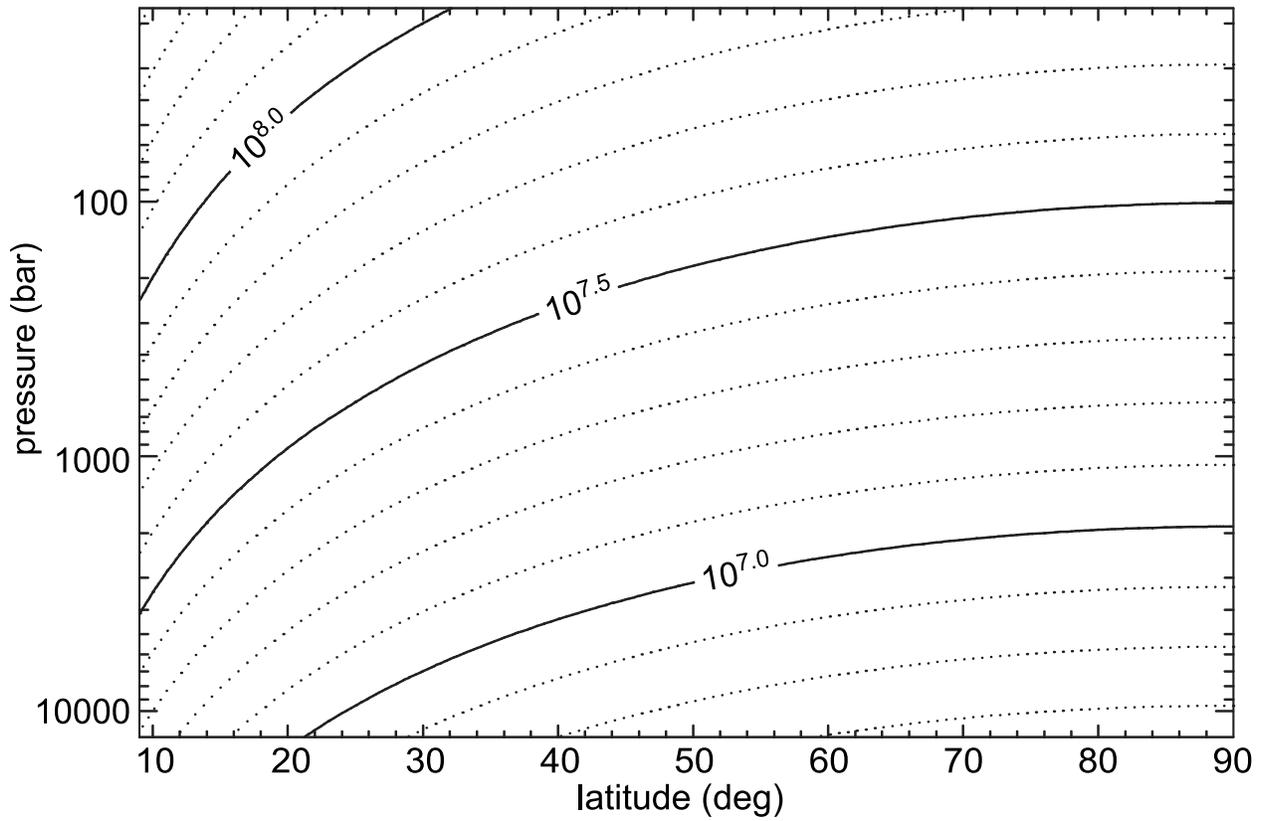}} \caption[Keddyrot latitude]{Eddy diffusion coefficient $K_{zz}$ at non-equatorial latitudes as a function of latitude and pressure predicted
using the theory of \citet{flasar1977,flasar1978} for thermally driven turbulent convection in a rapidly rotating system (see Eq.~(\ref{eq:Krot})).} \label{fig:Krotus}
\end{center}
\end{figure}

\begin{figure}
\begin{center}
{\includegraphics[angle=0,width=1.0\textwidth]{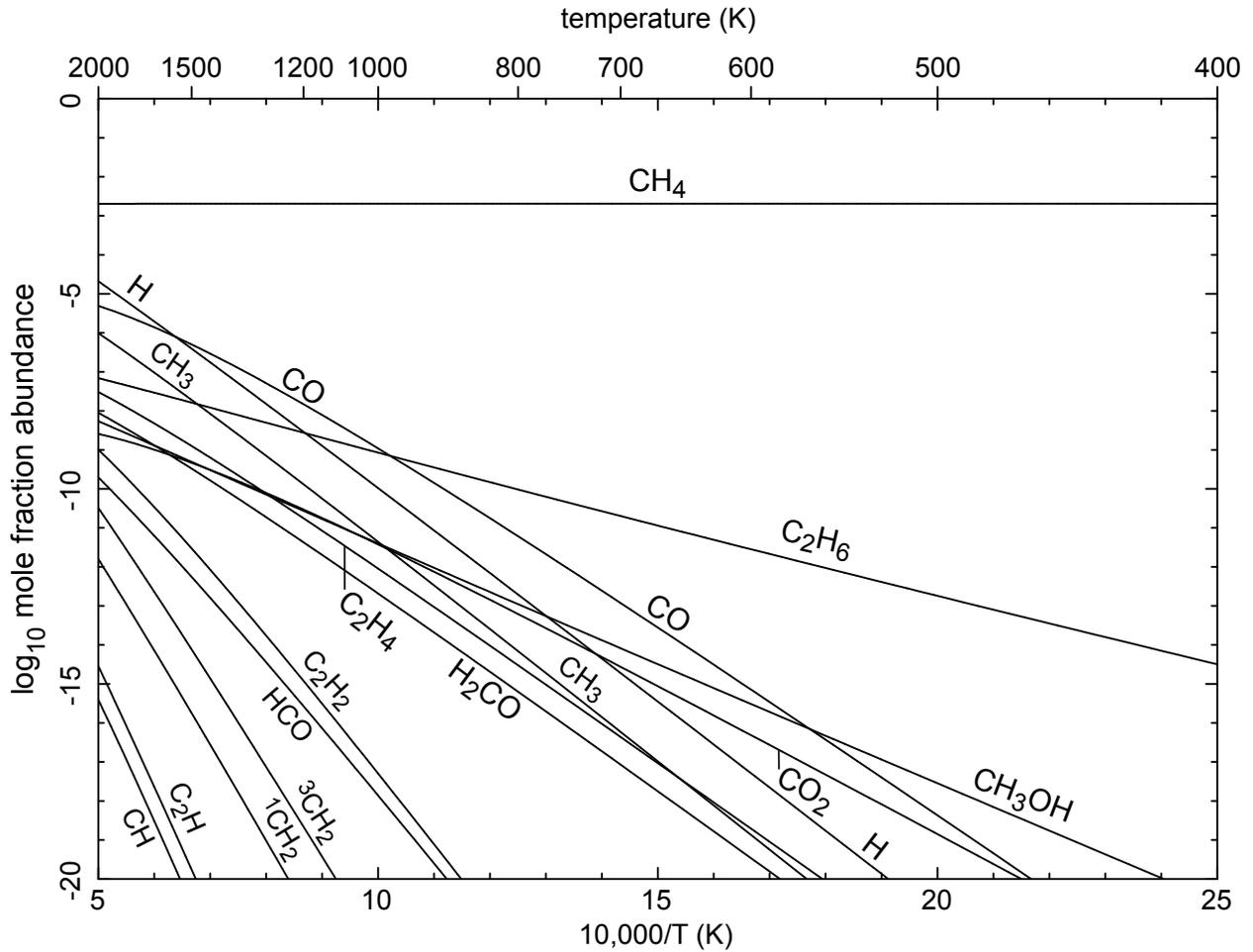}} \caption[Equilibrium Chemistry]{Carbon equilibrium chemistry for CH$_{4}$/H$_{2}$ = $2.37\times10^{{-3}}$ \citep{wong2004} and
H$_{2}$O/H$_{2}$ = $2.40\times10^{{-3}}$ (2.5x solar) in Jupiter's atmosphere, calculated using the NASA CEA code.  Modeled after Fig.~17 in \citet{fegley1994} for a slightly different
atmospheric composition and revised thermodynamic parameters for H$_{2}$CO and CH$_{2}$.} \label{fig:carbeq}
\end{center}
\end{figure}

\begin{figure}[p!]
\begin{center}
{\includegraphics[angle=0,width=1.0\textwidth]{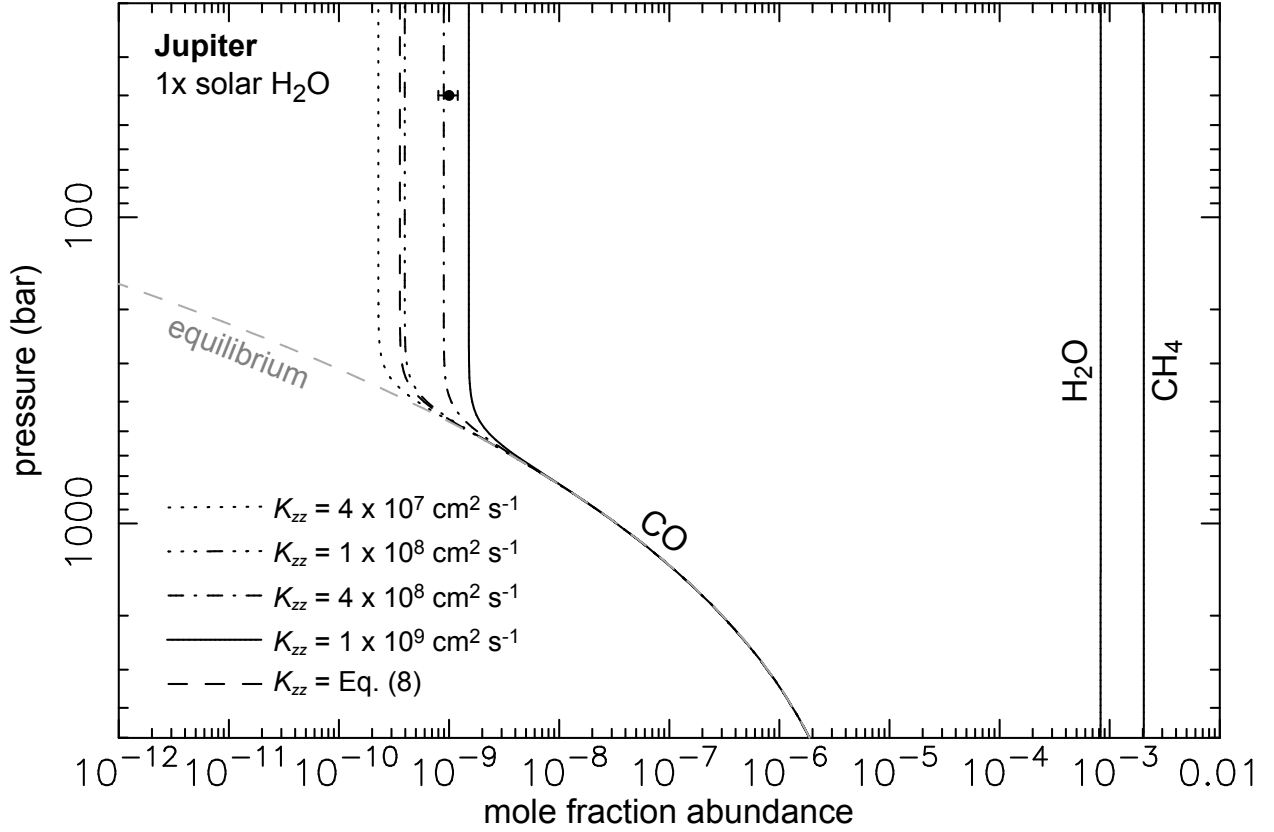}} \caption[CO Chemistry, Keddy]{Vertical profiles for the CO, H$_{2}$O, and CH$_{4}$ mole fractions in Jupiter's atmosphere for an assumed water
enrichment of 1x solar (where we assume the solar value is H$_{2}$O/H$_{2}=9.61\times10^{-4}$, \citet{lodders2004apj,lodders2009} and for assumed constant $K_{zz}$ values of $4\times10^{7}$
(dotted line), $1\times10^{8}$ (dot-dot-dot-dash line), $4\times10^{8}$ (dash-dot line), and $1\times10^{9}$ cm$^{2}$ s$^{-1}$ (solid line), as well as a $K_{zz}$ profile determined by
Eq.~(\ref{eq:Krot}) (dashed black line).  The dashed gray line represents the CO abundance predicted by chemical equilibrium using the NASA CEA code.  The circle with error bars represents the
observed tropospheric CO mole fraction reported by \citet{bezard2002}.  Note that if the deep water enrichment is 1x solar, as in the models shown here, the CO observations are best reproduced
for $K_{zz}$ = $4\times10^{8}$ cm$^2$ s$^{-1}$.  Note also that water and methane are the dominant oxygen- and carbon-bearing gases, respectively, throughout Jupiter's deep troposphere.}
\label{fig: CO keddy}
\end{center}
\end{figure}

\begin{figure}[p!]
\begin{center}
{\includegraphics[angle=0,width=1.0\textwidth]{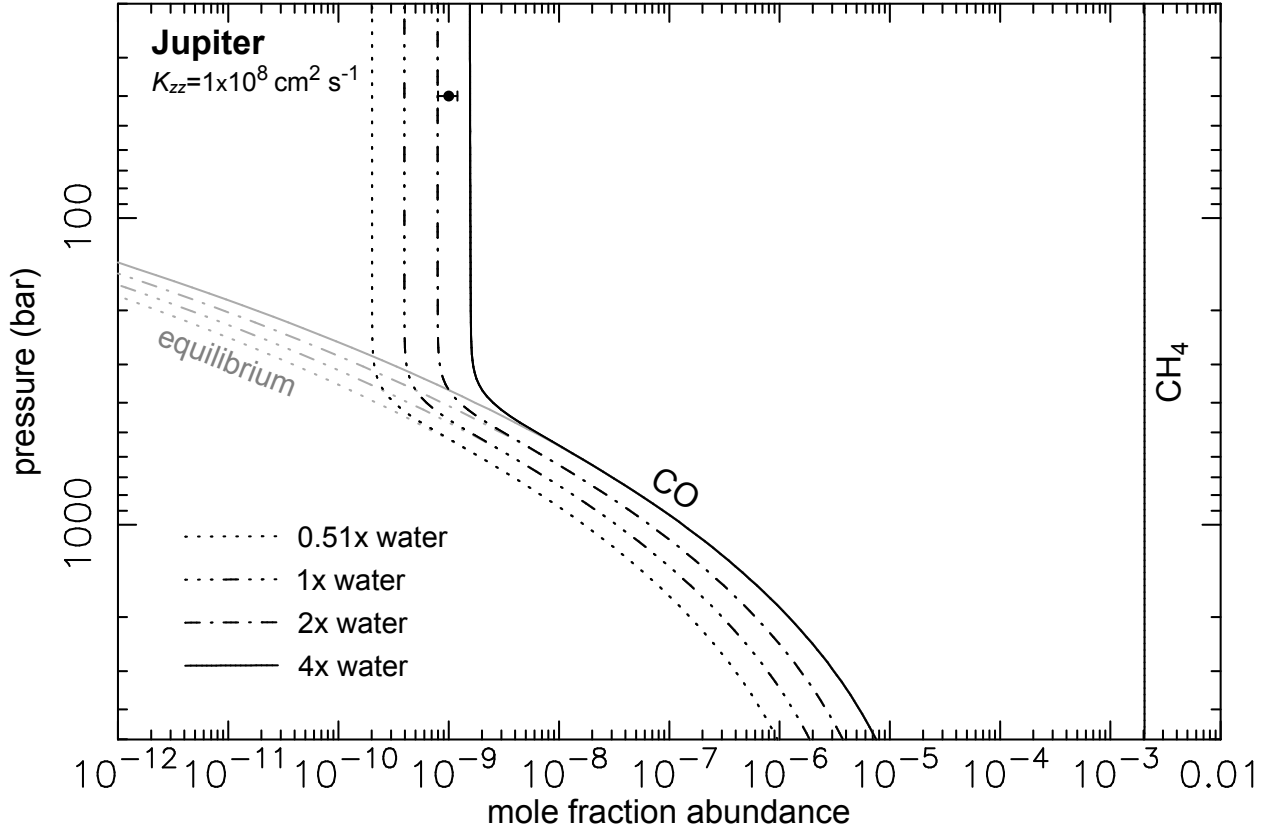}} \caption[CO Chemistry, Water]{Vertical profiles for CO and CH$_{4}$ in Jupiter's atmospere for an assumed eddy diffusion coefficient
$K_{zz}=1\times10^{8}$ cm$^{2}$ s$^{-1}$ (our nominal model) and for various assumed water enrichments ($E_{\textrm{H}_{2}\textrm{O}}$) of 0.51 (GPMS value; dotted line), 1 (dot-dot-dot-dash
line), 2 (dash-dot line) and 4 times (solid line) the solar H$_{2}$O/H$_{2}$ ratio of $9.61\times10^{-4}$ \citep{lodders2004apj,lodders2009}. The corresponding gray lines show the  CO abundance
predicted by chemical equilibrium using the NASA CEA code.  The circle with error bars represents the observed upper tropospheric CO mole fraction reported by \citet{bezard2002}.  Note that for
our assumed nominal value of $K_{zz}=1\times10^{8}$ cm$^{2}$ s$^{-1}$, the CO observations are best reproduced for assumed global water enrichments of 2-4 times solar.} \label{fig: CO water}
\end{center}
\end{figure}

\begin{figure}[p!]
\begin{center}
{\includegraphics[angle=0,width=1.0\textwidth]{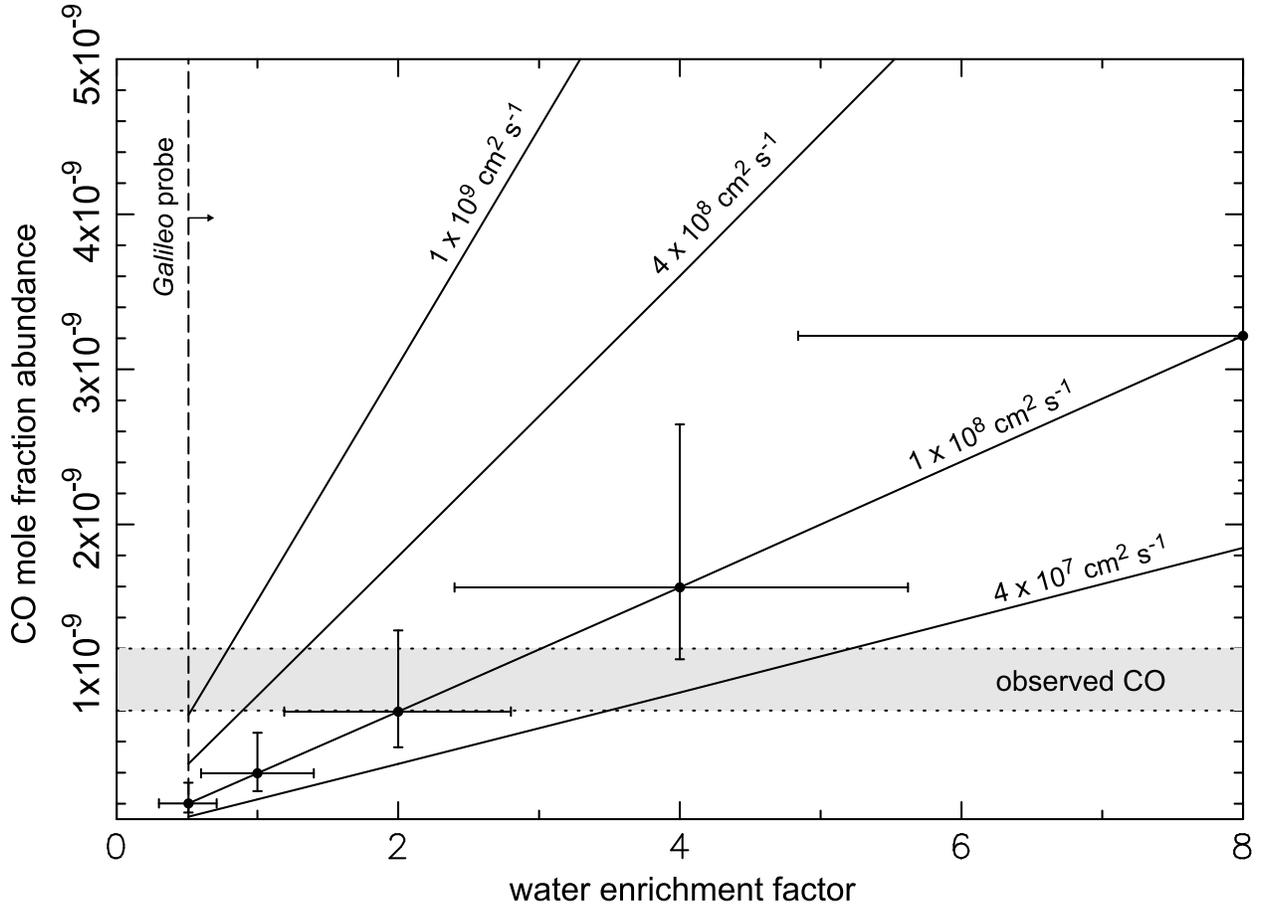}} \caption[Sensitivity to K and H2O]{The quenched CO mole fraction derived from our model (solid lines) as a function of the water enrichment
($E_{\textrm{H}_{2}\textrm{O}}$) relative to the solar value of H$_{2}$O/H$_{2}=9.61\times10^{-4}$ for different values of $K_{zz}$ (as labeled).  The dashed vertical line shows the water
abundance measured by the \textit{Galileo} entry probe \citep{wong2004}. The shaded area indicates the observed CO mole fraction of $(1.0\pm0.2)\times10^{-9}$ reported by \citet{bezard2002} and
constrains the range of $K_{zz}$ and $E_{\textrm{H}_{2}\textrm{O}}$ values in Jupiter's troposphere. The error bars for the $K_{zz}=1\times10^{8}$ cm$^{2}$ s$^{-1}$ solution represent the
estimated errors due to uncertainties in the reaction kinetics. Higher $K_{zz}$ values represent quenching and mixing from deeper in the atmosphere where CO is more abundant; higher water
enrichments yield higher CO abundances at the quench level.} \label{fig: CO sensitivity}
\end{center}
\end{figure}

\begin{figure}[p!]
\begin{center}
{\includegraphics[angle=0,width=1.0\textwidth]{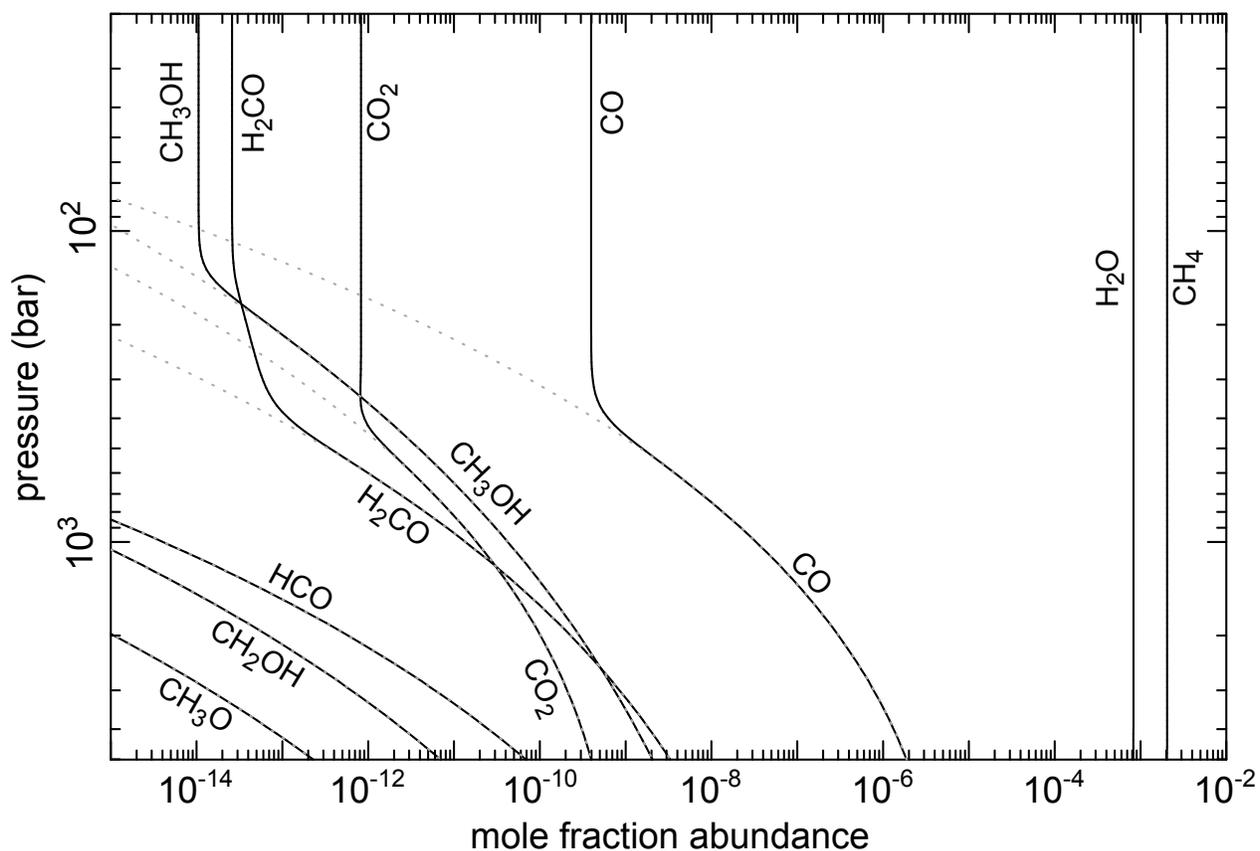}} \caption[Carbon-Oxygen Gases]{Vertical abundance profiles for oxidized carbon gases in Jupiter's atmosphere for $K_{zz}=1\times10^{8}$
cm$^{2}$ s$^{-1}$ and H$_{2}$O/H$_{2}=9.61\times10^{-4}$ (1x solar).  The dotted gray lines indicate thermochemical equilibrium; divergence from the equilibrium profiles (shown here for CO,
CO$_{2}$, H$_{2}$CO and CH$_{3}$OH) show where rapid vertical mixing and slow reaction kinetics drive each species to a constant quenched profile.  For HCO, CH$_{2}$OH, and CH$_{3}$O,
disequilibrium mixing occurs at much lower abundances (not shown).} \label{fig: carboxy gases}
\end{center}
\end{figure}

\begin{figure}[p!]
\begin{center}
{\includegraphics[angle=0,width=1.0\textwidth]{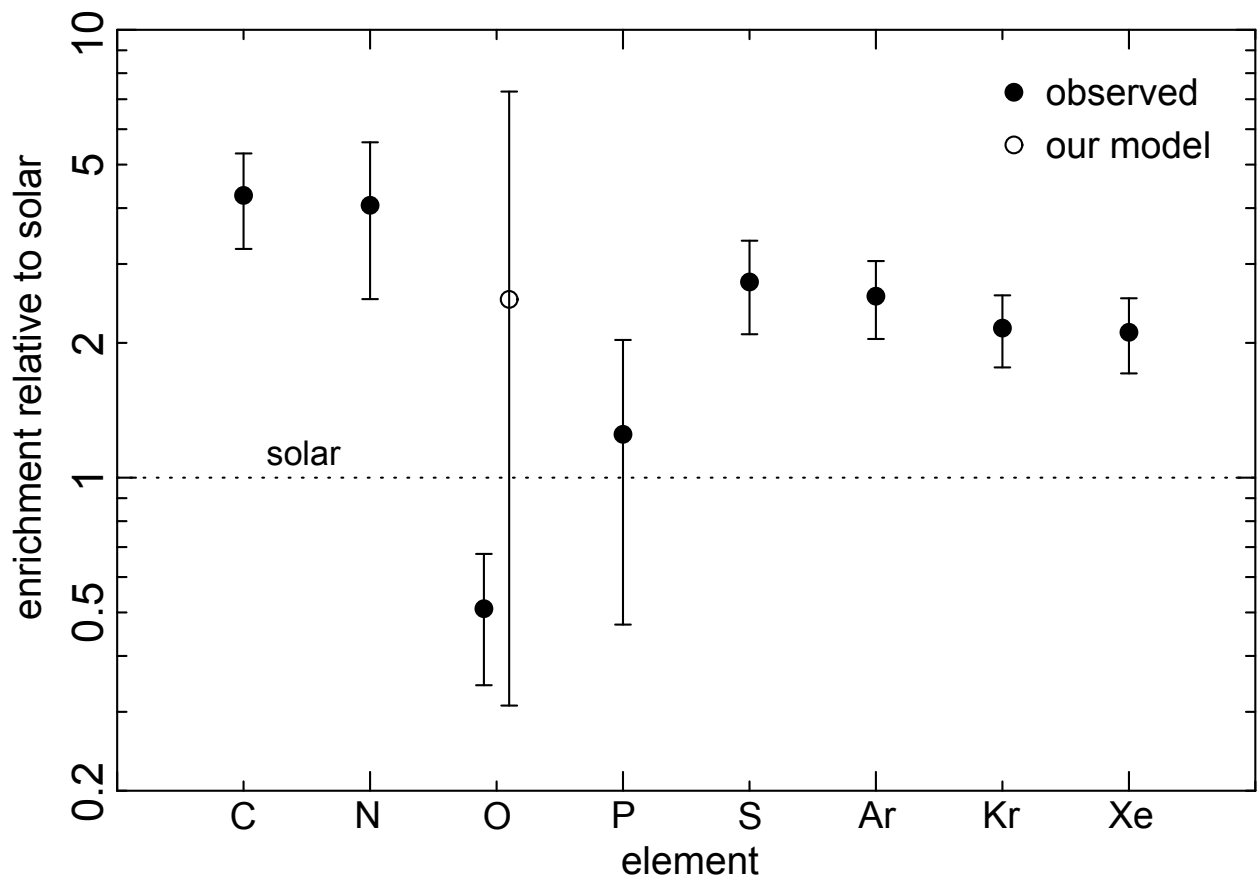}} \caption[Heavy Element Enrichments]{Observed Jovian heavy element enrichments relative to solar (filled circles), as measured by the
\textit{Galileo} entry probe \citep[C as CH$_{4}$, N as NH$_{3}$, O as H$_{2}$O, S as H$_{2}$S, Ar, Kr, Xe;][]{wong2004, mahaffy2000} and infrared spectroscopy (P as PH$_{3}$, see \S\ref{ss
Atmospheric Composition}), along with our kinetics-transport model results (open circle) for O as H$_{2}$O. Solar heavy element-to-hydrogen abundance ratios are taken from \citet{lodders2009}.
The water enrichments are defined relative to a solar water abundance of H$_{2}$O/H$_{2}=9.61\times10^{-4}$ (see text for details).} \label{fig: enrichments}\label{lastfigure}
\end{center}
\end{figure}

\end{document}